\documentclass[runningheads]{llncs}
\usepackage{hyperref}
\hypersetup{hidelinks}
\hypersetup{
	colorlinks   = true, 
}

\usepackage{graphicx}

\usepackage{tikz}
\usepackage{comment}
\usepackage{amsmath,amssymb} 
\usepackage{color}

\usepackage{multicol}
\usepackage{multirow}
\usepackage{booktabs}

\usepackage[accsupp]{axessibility}  
\usepackage[capitalize]{cleveref}
\usepackage{booktabs}
\usepackage[caption=false]{subfig}
\usepackage{bm}
\usepackage[linesnumbered,ruled]{algorithm2e}

\crefname{algocf}{Alg.}{Algs.}
\Crefname{algocf}{Algorithm}{Algorithms}

\definecolor{myorange}{RGB}{181, 115, 91}
\definecolor{myblue}{RGB}{125,161,177}
\definecolor{mygreen}{RGB}{151, 183, 150}
\definecolor{mygrey}{RGB}{156,156,156}
\newcommand{\drawpatternColor}[1]{
	$\left(\begin{tikzpicture}[x=1.5ex,y=1.5ex,baseline=0.2ex]\draw[color=black,semithick,fill=#1] (0,0) rectangle (1, 1);
	\end{tikzpicture}\right)$}

\begin{document}
\pagestyle{headings}
\mainmatter
\def\ECCVSubNumber{6046}

\title{Contextformer: A Transformer with Spatio-Channel Attention for Context Modeling in Learned Image Compression
} 

\titlerunning{Contextformer}
\author{A. Burakhan Koyuncu\inst{1,2} \and Han Gao\inst{4} \and Atanas Boev\inst{2} \and Georgii Gaikov\inst{3}\and \\ Elena Alshina\inst{2} \and Eckehard Steinbach\inst{1}}

\authorrunning{A. B. Koyuncu et al.}
\institute{Technical University of Munich, Munich, Germany\\ \email{burakhan.koyuncu@tum.de} \and Huawei Munich Research Center, Munich, Germany \and Huawei Moscow Research Center, Moscow, Russia \and Tencent America, Palo Alto, USA 	
}
\maketitle

\begin{abstract}
Entropy modeling is a key component for high-performance image compression algorithms. Recent developments in autoregressive context modeling helped learning-based methods to surpass their classical counterparts. However, the performance of those models can be further improved due to the underexploited spatio-channel dependencies in latent space, and the suboptimal implementation of context adaptivity. Inspired by the adaptive characteristics of the transformers, we propose a transformer-based context model, named Contextformer, which generalizes the de facto standard attention mechanism to spatio-channel attention. We replace the context model of a modern compression framework with the Contextformer and test it on the widely used Kodak, CLIC2020, and Tecnick image datasets. Our experimental results show that the proposed model provides up to 11\% rate savings compared to the standard Versatile Video Coding (VVC) Test Model (VTM)~16.2, and outperforms various learning-based models  in terms of PSNR and MS-SSIM.
\end{abstract}

\section{Introduction}
\label{sec:intro}
Recent works in learned image compression outperform hand-engineered classical algorithms such as JPEG~\cite{wallace1992jpeg} and BPG~\cite{bellard2015bpg}, and even reach the rate-distortion performance of recent versions of video coding standards, such as VVC~\cite{ohm2018versatile}. The most successful learning-based methods use an autoencoder based on~\cite{balle2018variational,minnen2018joint}, where the entropy of the latent elements is modeled and minimized jointly with an image distortion metric. The entropy modeling relies on two principles -- backward and forward adaptation~\cite{balle2020nonlinear}. The former employs a hyperprior estimator utilizing a signaled information source. The latter implements a context model, where previously decoded symbols are used for entropy estimation without a need for signaling. Due to its efficiency, a wide variety of context model architectures were explored in the recent literature ~\cite{guo2021causal,mentzer2018conditional,minnen2018joint,minnen2020channel,qian2021entroformer,qian2020learning,zhou2019multi}. We categorize those architectures into the following groups w.r.t. their targets: (1) increased support for spatial dependencies; (2) exploitation of cross-channel dependencies; (3) increased context-adaptivity in the entropy estimation. For instance, we consider the methods such as~\cite{cui2020g,cui2021asymmetric,burak2021,zhou2019multi} in the first category since those methods aim to capture long distant relations in the latent space. The 3D context~\cite{liu2019non,liu2019practical,mentzer2018conditional} and channel-wise autoregressive context model~\cite{minnen2020channel} fall in the second category. In those works, entropy estimation of each latent element can also use information from the spatially co-located elements of previously coded channels. In \cite{minnen2020channel} the authors show that entropy estimation, which mainly relies on cross-channel dependencies, outperforms their previous spatial-only model~\cite{minnen2018joint}. Most often, the context models use non-adaptive masked convolutions~\cite{li2021involution}. Those are location-agnostic~\cite{li2021involution}, i.e., the same kernel is applied to each latent position, which potentially reduces the model performance. Even for a larger kernel size, the performance return is marginal, as only a small set of spatial relations between symbols can be utilized. \cite{guo2021causal,qian2020learning} propose an adaptive context model, where the selection of latent elements to be used is based on pair-wise similarities between previously decoded elements. Furthermore, \cite{qian2021entroformer} uses a transformer-based context model to achieve context adaptivity for the spatial dimensions. However, those models have limited context adaptivity, as they are partially or not applying adaptive modeling for the cross-channel elements. For instance, in~\cite{guo2021causal}, although the primary channel carries on average $60\%$ of the information, the context model does not employ any adaptive mechanism for modeling it. 

Attention is a widely used deep learning technique that allows the network to focus on relevant parts of the input and suppress the unrelated ones~\cite{NIU202148}. In contrast to convolutional networks, attention-based models such as transformers \cite{vaswani2017attention} provide a large degree of input adaptivity due to their dynamic receptive field~\cite{naseer2021intriguing}. This makes them a promising candidate for a high-performance context model. Following the success of transformers in various computer vision tasks~\cite{carion2020end,dosovitskiy2020image,esser2021taming,jiang2021transgan,qian2021entroformer}, we propose a transformer-based context model, \textit{Contextformer}. Our contribution is threefold: (1) We propose a variant of the Contextformer, which adaptively exploits long-distance spatial relations in the latent tensor; (2) We extend the Contextformer towards a generalized context model, which can also capture cross-channel relations; (3) We present algorithmic methods to reduce the runtime of our model without requiring additional training.

In terms of PSNR, our model outperforms a variety of learning-based models, as well as VTM 16.2 \cite{jvet2019versatile} by a significant margin of 6.9\%\textendash10.5\% in average bits saving on the Kodak \cite{franzen1999kodak}, CLIC2020 \cite{CLIC2020} and Tecnick \cite{asuni2014testimages} image datasets. We also show that our model provides better performance than the previous works in a perceptual quality-based metric MS-SSIM \cite{wang2003scale}.
\section{Related Work}
\label{sec:background}
\subsection{Learned Image Compression}
Presently, the state-of-the-art in lossy image compression frameworks is fundamentally a combination of variational autoencoders and transform coding \cite{goyal2001theoretical}, where the classical linear transformations are replaced with learned non-linear transformation blocks, e.g., convolutional neural networks~\cite{balle2020nonlinear}. The encoder applies an analysis transform $g_a(\bm{x}; \bm{\phi})$ mapping the input image $\bm{x}$ to its latent representation $\bm{y}$. This transform serves as dimensionality reduction. The latent representation $\bm{y}$ is quantized by $Q(\cdot)$ and is encoded into the bitstream. In order to obtain the reconstructed image $\bm{\hat{x}}$, the decoder reads the quantized latent $\bm{\hat{y}}$ from the bitstream and applies the synthesis transform $g_s(\bm{\hat{y}};\bm{\theta})$, which is an approximate inverse of $g_a(\cdot)$.

Aiming to reduce the remaining coding redundancy in latent space, Ballé et al. \cite{balle2017end} introduced the factorized density model, which estimates the symbol distribution by using local histograms. During training, a joint optimization is applied to minimize both the symbol entropy and the distortion between the original and the reconstructed image. Knowledge of the probability distribution and coding methods such as arithmetic coding \cite{rissanen1979arithmetic} allows for efficient lossless compression of $\bm{\hat{y}}$. Later, Ballé et al.~\cite{balle2018variational} proposed using a hyperprior, which employs additional analysis and systhesis transforms $h_{a/s}(\cdot)$ and helps with modeling of the distribution $p_{\bm{\hat{y}}}(\bm{\hat{y}}|\bm{\hat{z}})$ conditioned on the side information $\bm{\hat{z}}$. The side information is modeled with a factorized density model, whereas $p_{\bm{\hat{y}}}(\bm{\hat{y}}|\bm{\hat{z}})$  is modeled as a Gaussian distribution. Their proposed framework can be formulated as
\begin{alignat}{2}
	&\bm{\hat{y}}= Q(g_a(\bm{x}; \bm{\phi})),\\
	&\bm{\hat{x}}= g_s(\bm{\hat{y}}; \bm{\theta})),\\
	&\bm{\hat{z}}= Q(h_a(\bm{\hat{y}}; \bm{\phi_h})),\\
	&p_{\bm{\hat{y}}}(\bm{\hat{y}}|\bm{\hat{z}})\leftarrow h_s(\bm{\hat{z}}; \bm{\theta_h}),
	\label{eq:1}
\end{alignat}
and the loss function $\mathcal{L}$ of end-to-end training is
\begin{alignat}{2}
	\mathcal{L}(\bm{\phi}, \bm{\theta}, \bm{\phi_h}, \bm{\theta_h},\bm{\psi}) &= \mathbf{R}(\bm{\hat{y}}) + \mathbf{R}(\bm{\hat{z}}) + \lambda \cdot \mathbf{D}(\bm{x}, \bm{\hat{x}})\\
	&= \mathbb{E}[\log_2(p_{\bm{\hat{y}}}(\bm{\hat{y}}|\bm{\hat{z}}))] + \mathbb{E}[\log_2(p_{\bm{\hat{z}}}(\bm{\hat{z}}|\bm{\psi}))] + \lambda \cdot \mathbf{D}(\bm{x}, \bm{\hat{x}}),
	\label{eq:2}
\end{alignat}
where $\bm{\phi}$, $\bm{\theta}$, $\bm{\phi_h}$ and $\bm{\theta_h}$ are the optimization parameters and $\bm{\psi}$ denotes the parameters of the factorized density model $p_{\bm{\hat{z}}}(\bm{\hat{z}}|\bm{\psi})$. $\lambda$ is the Lagrange multiplier regulating the trade-off between rate $\mathbf{R}(\cdot)$ and distortion $\mathbf{D}(\cdot)$.
\subsection{Context Model}
\label{sec:background2}
Higher compression performance requires more accurate entropy models, which would need an increased amount of side information~\cite{minnen2018joint}. To overcome this limitation, Minnen et al.~\cite{minnen2018joint} proposed a context model, which estimates the entropy of current latent element $\bm{\hat{y}_{i}}$ using the previously coded elements. Their approach extends \cref{eq:1} to
\begin{equation}
	p_{\bm{\hat{y}_{i}}}(\bm{\hat{y}_{i}}|\bm{\hat{z}})\leftarrow g_{ep}( g_{cm}(\bm{\hat{y}_{<i}}; \bm{\theta_{cm}}), h_s(\bm{\hat{z}}; \bm{\theta_h}); \bm{\theta_{ep}}),
\end{equation}
where the context model $g_{cm}(\cdot)$ is implemented as a 2D masked convolution. $g_{ep}(\cdot)$ computes the entropy parameters and $\bm{\hat{y}_{<i}}$ denotes the previously coded local neighbors of  current latent element $\bm{\hat{y}_{i}}$. Their proposed 2D context model requires 8.4\% fewer bits than BPG~\cite{bellard2015bpg}.

\begin{figure}[t]
	\centering
	\subfloat[\label{fig:sota_ctx_a}]{		\includegraphics[width=0.32\textwidth]{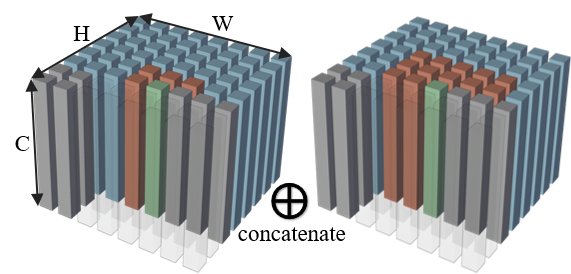}}
	\subfloat[\label{fig:sota_ctx_b}]{		\includegraphics[width=0.16\textwidth]{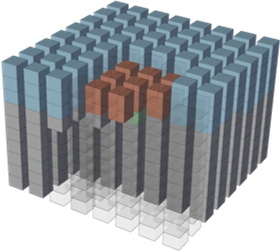}}
	\subfloat[\label{fig:sota_ctx_c}]{		\includegraphics[width=0.16\textwidth]{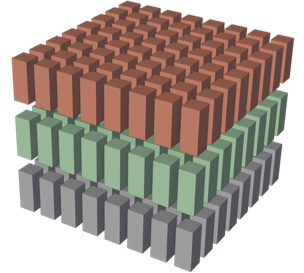}}
	\subfloat[ \label{fig:sota_ctx_d}]{		\includegraphics[width=0.16\textwidth]{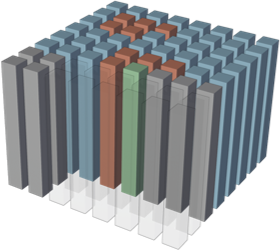}}
	\subfloat[\label{fig:sota_ctx_e}]{		\includegraphics[width=0.16\textwidth]{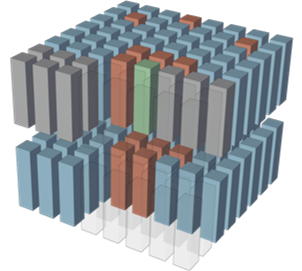}}
	
	\subfloat[\label{fig:our_ctx_a}]{		\includegraphics[width=0.15\textwidth]{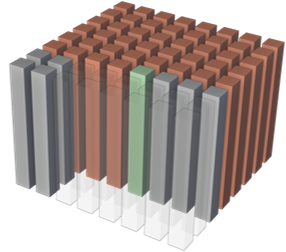}}
	\subfloat[\label{fig:our_ctx_b}]{		\includegraphics[width=0.135\textwidth]{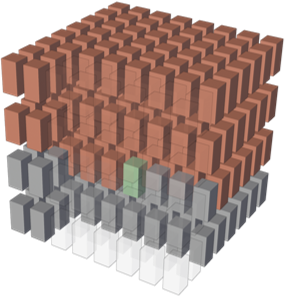}}
	\subfloat[\label{fig:our_ctx_c}]{		\includegraphics[width=0.135\textwidth]{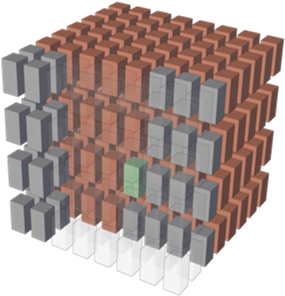}}
	\subfloat[\label{fig:our_ctx_d}]{		\includegraphics[width=0.18\textwidth]{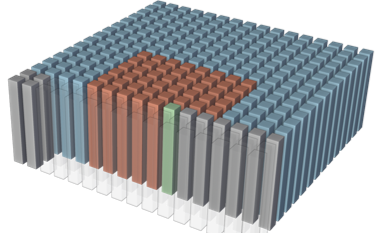}}
	\subfloat[\label{fig:our_ctx_e}]{		\includegraphics[width=0.17\textwidth]{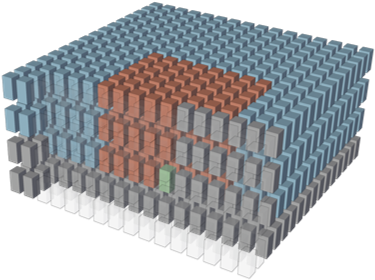}}
	\subfloat[\label{fig:our_ctx_f}]{		\includegraphics[width=0.17\textwidth]{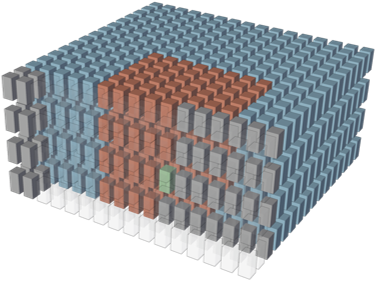}}
	\caption[]{
		Illustration of the latent elements used by the context model{\drawpatternColor{myorange}} to estimate the entropy of the current latent{\drawpatternColor{mygreen}} in (a\textendash e) for the prior-arts and (f\textendash k) our proposed context model. Previously coded and yet to be coded elements are displayed as{\drawpatternColor{myblue}} and{\drawpatternColor{mygrey}}, respectively. The displayed prior-art models are (a) multi-scale 2D context \cite{cui2020g,cui2021asymmetric,zhou2019multi}, (b) 3D context \cite{liu2019non,liu2019practical,mentzer2018conditional}, (c) channel-wise autoregressive context \cite{minnen2020channel}, (d) 2D context with global reference \cite{qian2020learning}, and (e) context with advanced global reference \cite{guo2021causal}. Note that in (c) each{\drawpatternColor{mygreen}} is coded simultaneously by using only a part of the elements presented as{\drawpatternColor{myorange}}, and in (e), the primary channel segment is shown at the bottom for better visibility. Our models with different configurations are shown in (f) Contextformer($N_{cs}{=}1$), (g) Contextformer($N_{cs}{>}1, sfo$), \mbox{(h) Contextformer($N_{cs}{>}1, cfo$)}. Note that the (serial) transformer-based context model of \cite{qian2021entroformer} employs similar mechanism as (f). (i\textendash k) show the versions of our models (f\textendash h)  using the sliding window attention.}
	\label{fig:all_ctx}
\end{figure}

Further improvements of the context model have been proposed (see  \cref{fig:sota_ctx_a,fig:sota_ctx_b,fig:sota_ctx_c,fig:sota_ctx_d,fig:sota_ctx_e}). In~\cite{cui2020g,cui2021asymmetric,zhou2019multi} a multi-scale context model was implemented, which employs multiple masked convolutions with different kernel sizes in order to learn various spatial dependencies simultaneously. \cite{liu2019non,liu2019practical,mentzer2018conditional} employ 3D masked convolutions in order to exploit cross-channel correlations jointly with the spatial ones.

Minnen and Singh proposed a channel-wise autoregressive context model~\cite{minnen2020channel}. It splits the channels of the latent tensor into segments and codes each segment sequentially with the help of a  previously coded segment. This reduces the sequential steps and outperforms the 2D context model of~\cite{minnen2018joint}. However, this approach uses only cross-channel correlations and omits the spatial ones.

Qian et al.~\cite{qian2020learning} proposed a context model, which combines 2D masked convolutions with template matching to increase the receptive field and provide context adaptivity. They search for a similar patch in the previously coded positions and use the best match as a global reference for the entropy model. 

Guo et al.~\cite{guo2021causal} proposed a context model, which can be seen as an extension of \cite{qian2020learning}. In their approach, the channels of the latent tensor are split into two segments. The first segment is coded with a 2D masked convolution, similar to~\cite{minnen2020channel}. Coding of the second segment is done using two different mechanisms: MaskConv+, an ``improved'' version of the 2D masked convolutions, and a global prediction. Additional to the local neighbors, MaskConv+ uses the spatially co-located elements from the first segment. The global prediction is made by calculating the similarity between all elements from the first segment. The indices of the top $k$ similar elements (from the corresponding position in the first segment) are used to select elements in the second segment and include those in the entropy model. They reported average bits savings of 5.1\% over VTM~8.0~\cite{jvet2019versatile}.

Qian et al.~\cite{qian2021entroformer} replaced the CNN-based hyperprior and context model with a transformer-based one which increased the adaptivity of the entropy model. They proposed two architectures for their context model -- serial and parallel model. The serial model processes the latent tensor sequentially similar to \cite{minnen2018joint}. The parallel one uses the checkboard like grouping prosed in \cite{he2021checkerboard} to increase the decoding speed. They achieved competitive performance with some of the CNN-based methods such as \cite{cheng2020learned}.
\subsection{Transformers}
Self-attention is an attention mechanism  originally proposed for natural language processing~\cite{vaswani2017attention}. Later, it was adopted in various computer vision tasks, where it outperformed its convolutional counterparts~\cite{carion2020end,dosovitskiy2020image,esser2021taming,jiang2021transgan}.

The general concept of a transformer is as follows. First, an input representation with $S$ sequential elements is mapped into three different representations, query \mbox{$\bm{Q} \in\mathbb{R}^{S\times d_q}$}, key $\bm{K}\in\mathbb{R}^{S\times d_k}$ and value $\bm{V}\in\mathbb{R}^{S\times d_v}$ with separate linear transformations. Then, the attention module dynamically routes the queries with the key-value pairs by applying a scaled dot-product. The attention module is followed by a point-wise multi-layer perceptron (MLP). Additionally, the multi-head attention splits the queries, keys, and values into $h$ sub-representations (so-called heads), and for each head, the attention is calculated separately. The final attention is computed by combining each sub-attention with a learned transformation ($\bm{W}$). The multi-head attention enables parallelization and each intermediate representation to build multiple relationships.

To preserve coding causality in autoregressive tasks, the attention mechanism has to be limited by a mask for subsequent elements not coded yet. The masked multi-head attention can be formulated as:
\begin{alignat}{2}
	&Attn(\bm{Q},\bm{K},\bm{V})= \text{concat}({head}_1,\ldots,{head}_h)\bm{W},\\
	&head_i(\bm{Q}_i,\bm{K}_i,\bm{V}_i)=\text{softmax}\left(\frac{\bm{Q}_i\bm{K}_i^T}{d_k}\odot\bm{M}\right)\bm{V}_i,
	\label{eq:multihead}
\end{alignat}
where $\bm{Q}_i\in\mathbb{R}^{S\times \frac{d_q}{h}}$, $\bm{K}_i\in\mathbb{R}^{S\times \frac{d_k}{h}}$ and $\bm{V}_i\in\mathbb{R}^{S\times\frac{d_v}{h}}$ are the sub-representations, the mask $\bm{M} \in\mathbb{R}^{S\times S}$ has ones in its lower triangle and the rest of its values are minus infinity, and $\odot$ stands for the Hadamard product.

\section{Our Approach}
\label{sec:proposed}
In this section, we introduce our transformer-based context model, \textit{Contextformer}, which provides context adaptivity and utilizes distant spatial relations in the latent tensor. We present two versions of the model: a simple version, which uses spatial attention; and a more advanced version, which employs attention both in the spatial and channel dimensions for entropy modeling.
\begin{figure}[t]
	\centering
	
	\tabskip=0pt
	\valign{#\cr
		\hbox{%
			\subfloat[\label{fig:full_model_a}]{\includegraphics[trim=3.45cm 0cm 3.71cm 0cm, clip,width=0.56\textwidth]{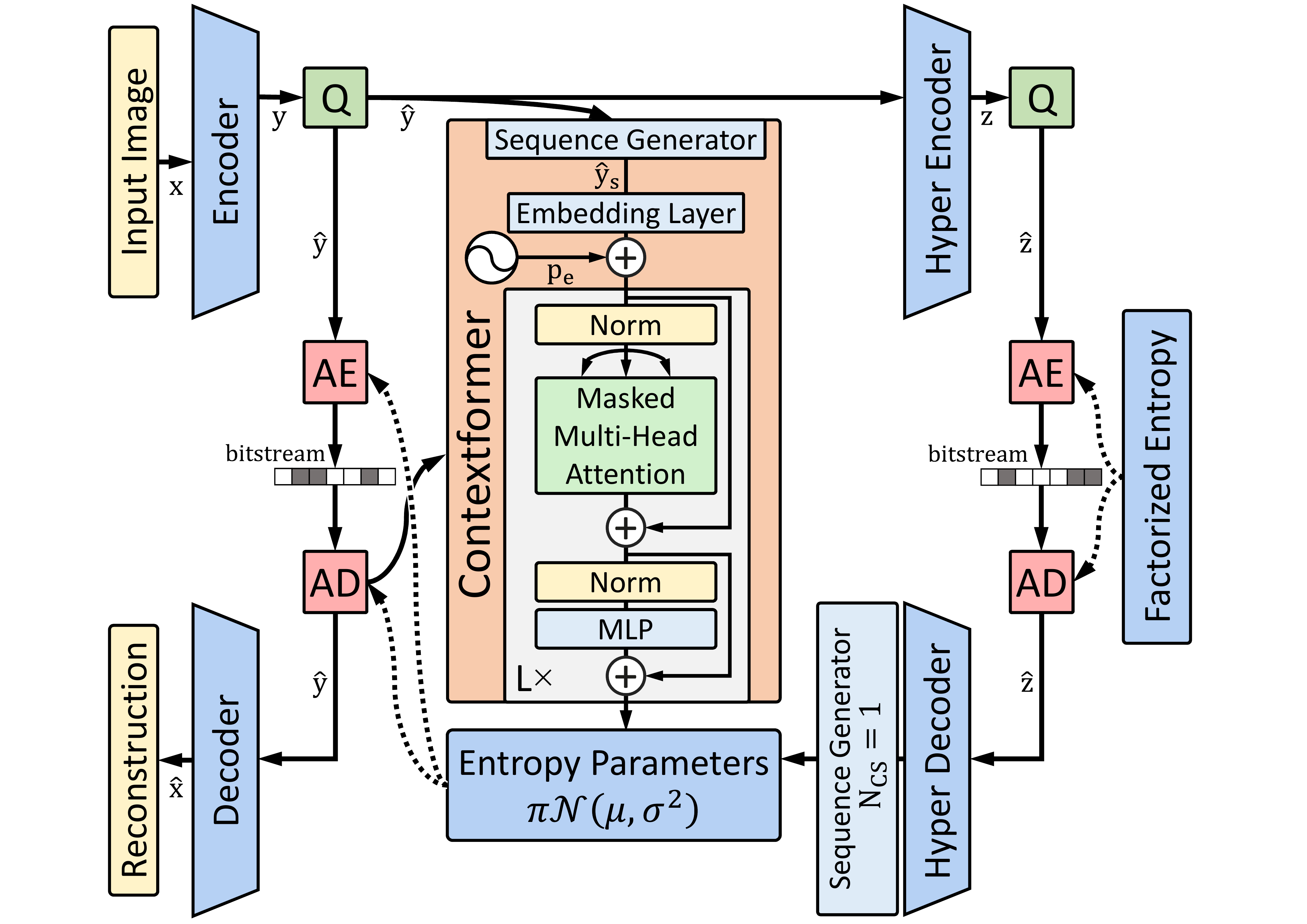}}
		}\cr
		\noalign{\hfill}
		\hbox{%
			\subfloat[\label{fig:full_model_b}]{\includegraphics[trim=2.88cm 9.1cm 2.99cm 1.253cm, clip,width=0.40\textwidth]{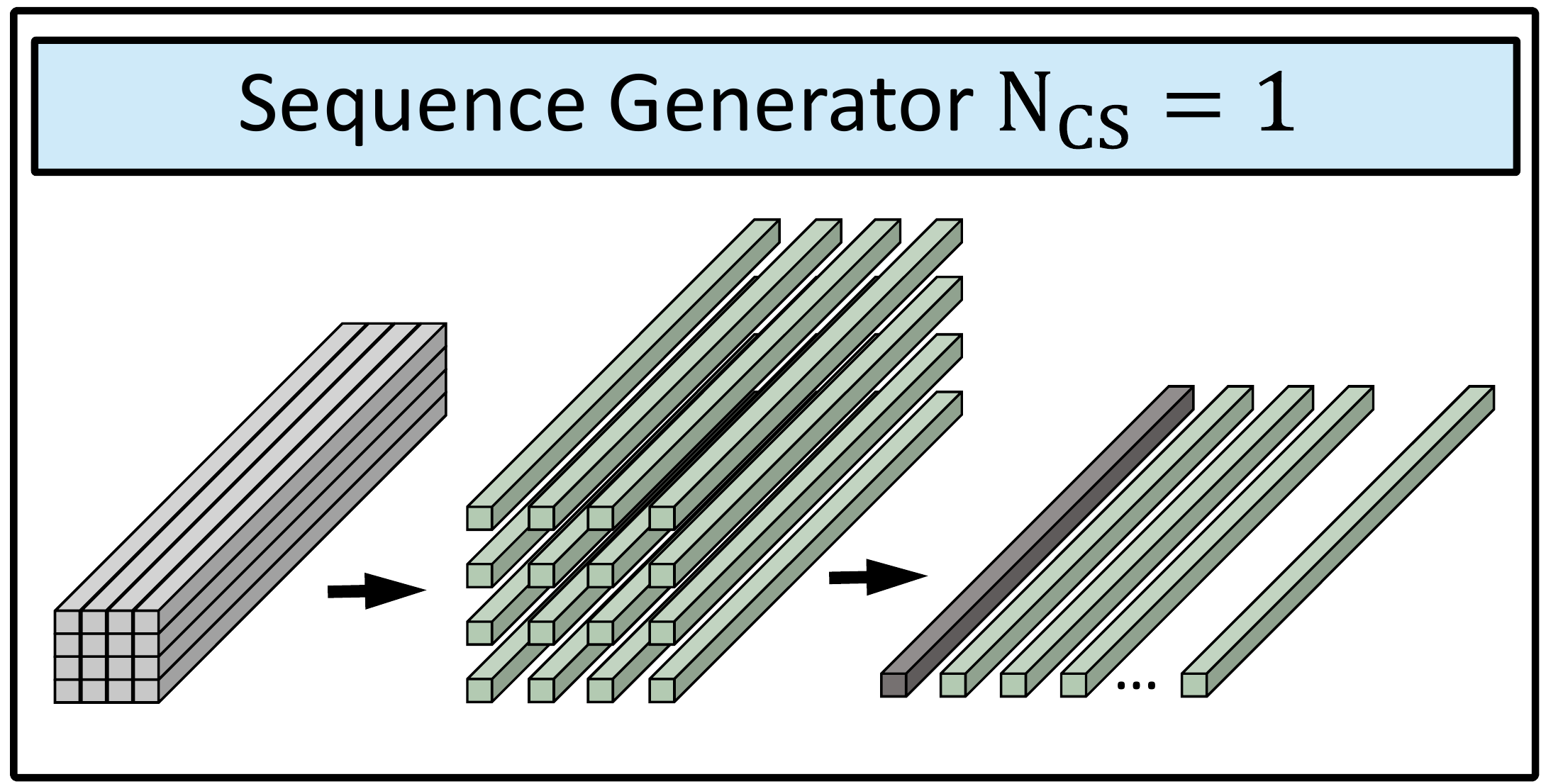}}	
		}\vfill
		\hbox{%
			\subfloat[\label{fig:full_model_c}]{\includegraphics[trim=2.88cm 9.1cm 2.99cm 1.253cm, clip,width=0.40\textwidth]{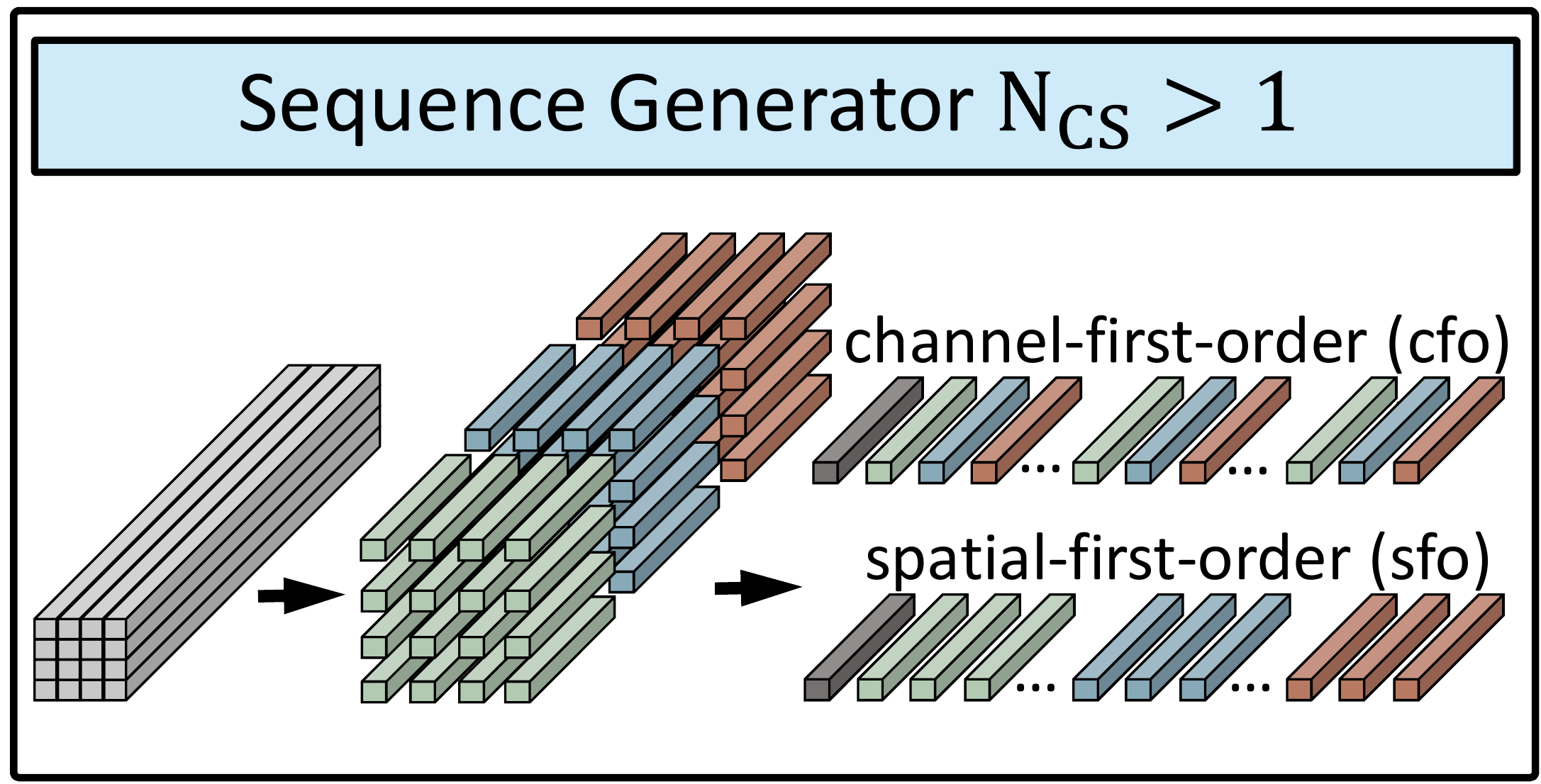}}
		}\cr
	}
	\caption{Illustration of (a) our proposed model with the Contextformer, (b) sequence generator for the Contextformer with spatial attention, Contextformer$(N_{cs}{=}1)$, and (c) sequence generator for the Contextformer with spatio-channel attention, Contextformer$(N_{cs}{>}1)$. The prepended start token is shown in dark gray in (b-c). Inspired by \cite{lee2018context}, we use channel-wise local hyperprior neighbors to increase performance; thus, regardless of the selected $N_{cs}$, we apply the sequence generator depicted in (b) to the output of the hyperdecoder.}
\end{figure}
\subsection{Contextformer with Spatial Attention}
\label{sec:contextformer}
The proposed Contextformer builds on top of the architecture introduced in~\cite{cui2021asymmetric}. In the encoder, this model employs $3{\times}3$ convolution layers with GDN activation function~\cite{balle2015density} and residual non-local attention modules (RNAB)~\cite{zhang2018residual}. The structure of the decoder is very similar to the one of the encoder, with the exception that residual blocks (ResBlock)~\cite{liu2019non} are used in the first layer to enlarge the receptive field. Additionally, the model adopts a hyperprior network, the multi-scale context model~\cite{zhou2019multi} and the universal quantization~\cite{ziv1985universal}. This model estimates the distribution $p_{\bm{\hat{y}}}(\bm{\hat{y}}|\bm{\hat{z}})$ with a single Gaussian distribution. In our approach, we use a Gaussian mixture model~\cite{cheng2020learned} with 3 mixture components $k_m$, which is known to increase the accuracy of the entropy model.

In contrast to Cui et al.~\cite{cui2021asymmetric}, we use a Contextformer instead of their multi-scale context model, as shown in \cref{fig:full_model_a}. First, the latent $\bm{\hat{y}}\in\mathbb{R}^{H \times W \times C}$ is rearranged into a sequence of spatial patches $\bm{\hat{y}}_s\in\mathbb{R}^{\frac{HW}{p_h p_w} \times (p_h p_w C)}$. Here, $H$, $W$ and $C$ stand for the height, width and number of channels; $(p_h,p_w)$ corresponds to the shape of each patch. Usually, patch-wise processing reduces complexity, especially for large images~\cite{dosovitskiy2020image}. However, the latent $\bm{\hat{y}}$ already has a 16 times lower resolution than the input image, which makes learning an efficient context model harder and leads to a performance drop. To remedy this issue, we set the patch size to $1\,{\times}\,1$, so each sequential element corresponds to one pixel in the latent tensor (see \cref{fig:full_model_b}).

The Contextformer has $L$ transformer layers with a similar architecture to that of ViT~\cite{dosovitskiy2020image}. Each layer requires an intermediate tensor with an embedding size of $d_e$. Therefore, we apply a learnable linear transformation $\mathbb{R}^{HW \times C} \rightarrow \mathbb{R}^{HW \times d_e}$ (embedding layer). In order to introduce permutation-variance, we add a learned position encoding similar to the one in~\cite{dosovitskiy2020image,esser2021taming}, but we apply it to the first layer only. We prepend the latent sequence $\bm{\hat{y}_s}$ with a zero-valued start token to ensure the causality of coding. We use masking in the attention as described in \cref{eq:multihead}, and multi-head attention with 12 heads. Multi-head allows our model to independently focus on different channel segments of $\bm{\hat{y}}$.

\subsection{Contextformer with Spatio-Channel Attention}
Although multi-head attention is computationally efficient in handling cross-channel dependencies, it can explore relationships in a single channel only partially. For example, consider a Contextformer with a single transformer layer. Given a latent tensor $\bm{\hat{y}}\in\mathbb{R}^{H \times W \times C}$ and its sequential representation $\bm{\hat{y}}_s \in\mathbb{R}^{S\times C}$; the $n$-th sub-representation of the sequence $\bm{\hat{y}}_s(n,h_i)\in\mathbb{R}^{1\times\frac{C}{h}}$ can only attend to the previous representations $\bm{\hat{y}}_s({<}n,h_i)$ with the same head index $h_i$. This means that the attention between different channel segments is not considered. Another limitation arises from the way the model behaves w.r.t. $\bm{\hat{y}}$. For modeling the entropy of latent element $\bm{\hat{y}}(i,j,c)$ (with spatial coordinates $(i,j)$ and channel index $c$), the context model cannot access information from the spatially co-located elements from other channels $\bm{\hat{y}}(i,j,{\neq}c)$. This limits exploiting the cross-channel dependencies in $\bm{\hat{y}}$, and, therefore, the performance of the model.

To remedy this issue, we generate spatio-channel patches in the latent space $\bm{\hat{y}}_s\in\mathbb{R}^{\frac{HWC}{p_h p_w p_c} \times (p_h p_w p_c)}$, where  $p_c$ corresponds to the size of each channel segment, and total number of channel segments is $N_{cs}\,{=}\,\frac{C}{p_c}$. In a special case of $(p_h,p_w)\,{=}\,(H,W)$, our patch generation method is similar to the slicing method in channel-wise context modeling~\cite{minnen2020channel}, but our model has a multi-head attention added. In this work, we set $p_h$ and $p_w$ to 1 as already discussed in~\cref{sec:contextformer}. Splitting the latent tensor into multiple channel segments enables two different coding methods, spatial-first-order ($sfo$) and channel-first-order ($cfo$). The first method prioritizes the spatial dimensions and codes all latent elements from a channel segment sequentially before starting with the next segment. The second method prioritizes the channel dimension, and codes all channel segments with the same spatial coordinate sequentially, before coding elements from the next spatial coordinate.

Spatio-channel sequence generation allows the standard transformer to use channel attention, from which a generalized context model can be obtained. To illustrate this, we compare how dependencies in the latent space are handled by Contextformer for various $N_{cs}$, and how those dependencies are handled by context models in the prior-art.

\subsubsection{In case of $N_{cs}\,{=}\,1$.} The attention is limited to a spatial one (see \cref{sec:contextformer}). However, such a model provides faster encoding and decoding due to the smaller number of required autoregressive steps and still has better performance than some models in the prior-arts. As illustrated in \cref{fig:our_ctx_a}, Contextformer\,$(N_{cs}\,{=}\,1)$ has a larger receptive field than~\cite{cui2020g,cui2021asymmetric,zhou2019multi}, and also employs learned context adaptivity. Additionally, in \cite{qian2020learning} the receptive field is limited to a single reference and its neighboring latent elements, whereas Contextformer\,$(N_{cs}\,{=}\,1)$'s receptive field is dynamic and theoretically unlimited. Notably, the serial model of \cite{qian2021entroformer} (best performing one) uses a similar context model as Contextformer\,$(N_{cs}\,{=}\,1)$. However, their model has only a sparse-attention mechanism, where\-as Contextformer employs the full attention mechanism.

\subsubsection{In case of $C\,{\geqq}\,N_{cs}\,{>}\,1$.} We achieve a context model that can exploit both spatial and cross-channel dependencies. As shown in~\cref{fig:our_ctx_b,fig:our_ctx_c}, both Contextformers\,$(N_{cs}\,{>}\,1)$ with different coding order handle spatio-channel relationships. Moreover, in \cite{guo2021causal} only non-primary channel segments could be selected for entropy estimation, while the receptive field of the Contextformer's receptive field can adapt to cover every channel segment. Compared to \cite{qian2021entroformer}, the Contextformer\,$(N_{cs}\,{>}\,1)$ provides a more adaptive context model due to the spatio-channel attention. In the extreme case $N_{cs}\,{=}\,C$, the model employs the spatio-channel attention to its full extend by computing the attention between every single latent element. Other implementations can be seen as a simplification of the extreme case for balancing the trade-off between performance and complexity. The Contextformer\,$(N_{cs}\,{=}\,C)$ can be regarded as a 3D context model with a large and adaptive receptive field. 

\subsection{Handling High-Resolution Images}
Although the receptive field size of the Contextformer is theoretically unlimited, computing attention for long input sequences, e.g., high resolution images, is not feasible due to the quadratic increase of memory requirement and computational complexity with increasing the length of the input sequence. Therefore, we have to limit the receptive field of our model and use sliding-window attention as described in~\cite{esser2021taming,parmar2018image}. Inspired by 3D convolutions, we implemented a 3D sliding-window to traverse the spatio-channel array. Unlike 3D convolutions, the receptive field only slides across spatial dimensions and expands to encompass all elements in the channel dimension. In \cref{fig:our_ctx_d,fig:our_ctx_e,fig:our_ctx_f} one can see the sliding-window attention mechanism for various Contextformer variants. For computational efficiency, during training, we used fixed-sized image patches and omitted the sliding-window operations. During inference, we set the size of the receptive field according to the sequence length ($HWN_{cs}$) used for training.

\subsection{Runtime optimization}
Generally, autoregressive processes cannot be efficiently implemented on a GPU due to their serialized nature~\cite{he2021checkerboard,burak2021,minnen2020channel}. One commonly used approach~\cite{devlin-etal-2019-bert,vaswani2017attention} is to pad a set of sequences to have a fixed length, thus enabling the processing of multiple sequences in parallel during training (we refer to this method as Pad\&Batch). A straightforward implementation of the sliding-window attention uses dynamic sequence lengths for every position of the window. We call this \textit{dynamic sequence} processing (DS). The padding technique can be combined with the sliding-window by applying masking to the attention mechanism. However, one still needs to calculate attention for each padded element, which creates a bottleneck for the batched processing.

We propose a more efficient algorithm to parallelize the sliding-window attention. The first step of the algorithm calculates the processing order (or \textit{priority}) for every position of the sliding window and then groups the positions with the same processing order for batch processing. One possible processing order is to follow the number of elements in each window and process them from the smallest to the largest number of elements. We refer to this method as Batched Dynamic Sequence (BDS). Note that transformers are sequence-to-sequence models; they simultaneously compute an output for each element in a sliding window and preserve causality of the sequence due to the masking. Therefore, we can skip computation of intermediate channel segments and calculate the output of the last channel segment for each spatial coordinate of the sliding window, which we refer to as skipping intermediate channel segments (SCS). It is worth mentioning that both the BDS and SCS methods can only be applied in the encoder, where all elements of the latent tensor are simultaneously accessible. For the decoder-side runtime optimization, we adopted the wavefront coding described in~\cite{9067005}, which is similar to the partitioning slices used in VVC~\cite{articleewr}. We use the same processing priority to the independent sliding windows along the same diagonal, which allows for simultaneous decoding of those windows. More information about the proposed runtime optimization algorithms can be found in the supplementary materials.
\section{Implementation Details}
\label{sec:impl}
We present a few variants of the Contextformer by changing its parameters $\{L,\,d_e,\,N_{cs},\,co\}$, where $L$, $d_e$ and $N_{cs}$ correspond to number of layers, embedding size and number of channel segments, and $co$ corresponds to the coding order -- either spatial-first ($sfo$) or channel-first ($cfo$). For all models, we used the same number of heads $h$ and MLP size of $\mathit{d}_{mlp}$. We selected the base configuration of the Contextformer as $\{L\,{=}\,8,\,d_e\,{=}\,384,\,d_{mlp}\,{=}\,4d_e,\,h\,{=}\,12,\,N_{cs}\,{=}\,4,\,\allowbreak co \,{=}\, cfo\}$. More information about the architectural details can be found in the suplementary materials.

For training of all variants, we used random $256\,{\times}\,256$ image crops from the \mbox{Vimeo-90K} dataset \cite{xue2019video}, batch size of 16, and ADAM optimizer \cite{kingma2014adam} with the default settings. We trained our models for 120 epochs (${\sim}$1.2M iterations). Following \cite{begaint2020compressai}, we used the initial learning rate of $10^{-4}$ and reduced it by half every time the validation loss is nearly constant for 20 epochs. For this purpose, we used the validation set of Vimeo-90K. We selected mean-squared-error (MSE) as the distortion metric $\mathbf{D}(\cdot)$ and trained a set of models with $\lambda\in \{0.002,\,0.004,\,0.007,\,0,014,\,0.026,\,0.034,\,0.070\}$ to cover various bitrates. We also obtained models optimized for MS-SSIM \cite{wang2003multiscale} by finetuning the models trained for MSE, in the same fashion for ${\sim}$500K iterations. By default, we selected both intermediate layer size $N$ of the encoder and decoder, and bottleneck size $M$ as 192. To increase the model capacity at the high target bitrates ($\lambda_{5,6,7}$), we increased $M$ from 192 to 312 by following the common practice \cite{cheng2020learned,minnen2018joint,minnen2020channel}.
\section{Experimental Results}
\label{sec:ex}
Unless specified otherwise, we used the base configuration (see \cref{sec:impl}) and tested its performance on the Kodak image dataset \cite{franzen1999kodak}. We set the spatial receptive field size to $16{\times}16$ for the sliding-window attention. We compared the performance with the following models: the 2D context models by Minnen et al.~\cite{minnen2018joint} and Cheng et al.~\cite{cheng2020learned}, the multi-scale 2D context model by Cui et al.~\cite{cui2021asymmetric}, the channel-wise autoregressive context model by Minnen and Singh~\cite{minnen2020channel}, the context model with an advanced global reference by Guo et al.~\cite{guo2021causal}, and the transformer-based context model by Qian et al. \cite{qian2021entroformer}. When the source was available, we ran the inference algorithms of those methods; in other cases, we took the results from the corresponding publications. For a fair comparison, we used the model from \cite{guo2021causal} without the GRDN post-filter \cite{kim2019grdn}. Similarly, we used the serial model from \cite{qian2021entroformer}, since it performs better and is more related to our approach. We also compared with the results achieved by classical compression algorithms such as BPG~\cite{bellard2015bpg} and VTM~16.2~\cite{jvet2019versatile}. In order to test the generalization capability of our model, we also tested its performance on CLIC2020 \cite{CLIC2020}, and Tecnick \cite{asuni2014testimages} datasets. Additionally, we present the impact of different configurations of our model on its complexity. More details about the performance comparison, and examples of compression artifacts can be found in the supplementary materials.

\begin{figure}
	\centering
	\subfloat[\label{fig:res1}]{\includegraphics[width=0.495\linewidth]{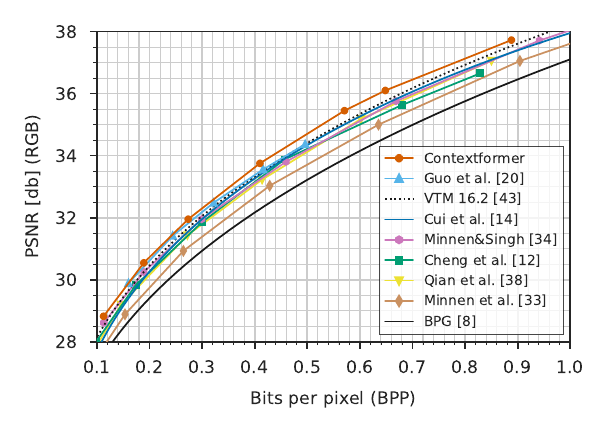}}
	\subfloat[\label{fig:res2}]{\includegraphics[width=0.495\linewidth]{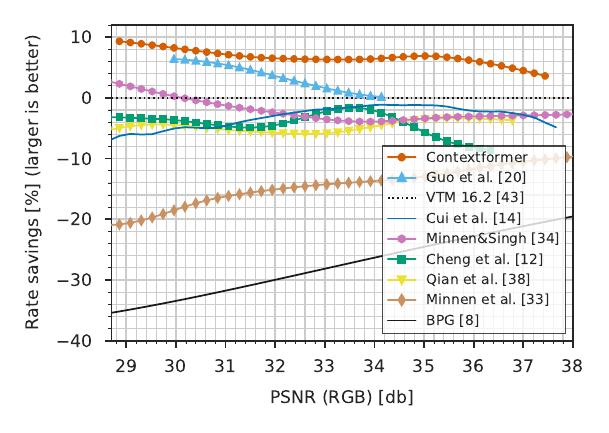}}
	\caption{Illustration of (a) the rate-distortion performance and (b) the rate savings relative to VTM 16.2 as a function of PSNR on the Kodak dataset showing the performance of our model compared to various learning-based and classical codecs.}
	\label{fig:res12}
\end{figure}

\subsubsection{Performance.}
\label{sec:performance}
In \cref{fig:res1} we show the rate-distortion performance of the Contextformer with a spatio-channel attention mechanism and the comparative performance of prior methods on the Kodak image dateset. Our method qualitatively surpasses the rest in terms of PSNR for all rate points under test. According to the Bjøntegaard Delta rate (BD-Rate) \cite{bjontegaard2001calculation}, our method achieves average saving of $6.9\%$ over VTM 16.2, while the model in~\cite{guo2021causal} provides $3.7\%$ saving over the same baseline. On average, our model saves $10\%$ more bits compared to the multi-scale 2D context model in~\cite{cui2021asymmetric}. Notably, the only difference between our model and the one in~\cite{cui2021asymmetric} is the context and entropy modeling, and both methods have similar model sizes. The performance of our method and the prior in terms of the generalized BD-Rate metric~\cite{minnen2020channel} is shown in \cref{fig:res2}. Our model achieves state-of-the-art performance by reaching $9.3\%$ rate savings for low bitrate and $4\%$ rate savings at the highest quality over VTM 16.2.

\begin{figure}
	\centering
	\includegraphics[width=0.5\linewidth]{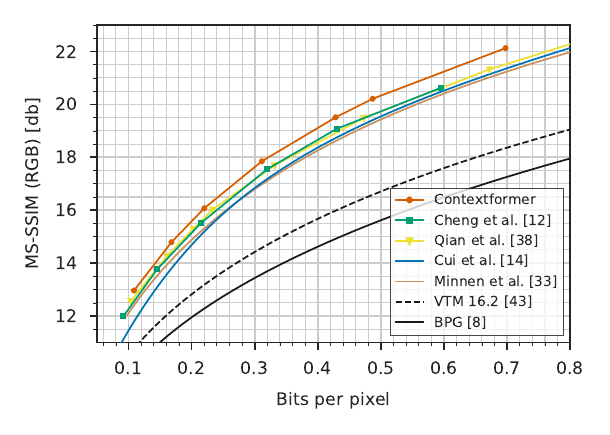}
	\caption{Illustration of the rate-distortion performance in terms of MS-SSIM on Kodak dataset showing the performance of our model compared to various learning-based and classical codecs. All learned methods were optimized for MS-SSIM.}
	\label{fig:res8}
\end{figure}

We also evaluated our model optimized for MS-SSIM \cite{wang2003multiscale}. \cref{fig:res8} shows that our model also outperforms previous methods for this perceptual quality metric. On average, our models saves $8.7\%$ more bits than Cheng et al.~\cite{cheng2020learned}.

\subsubsection{Generalization Capability.}
\begin{figure}[t]
	\centering
	\subfloat[\label{fig:res3}]{\includegraphics[width=0.495\linewidth]{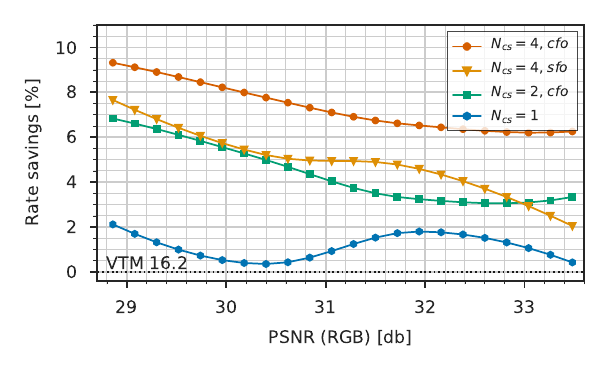}}
	\subfloat[\label{fig:res4}]{\includegraphics[width=0.495\linewidth]{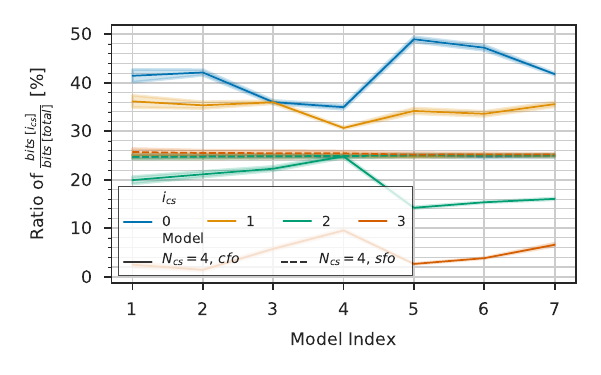}}
	\hfill
	\subfloat[\label{fig:res5}]{\includegraphics[width=0.495\linewidth]{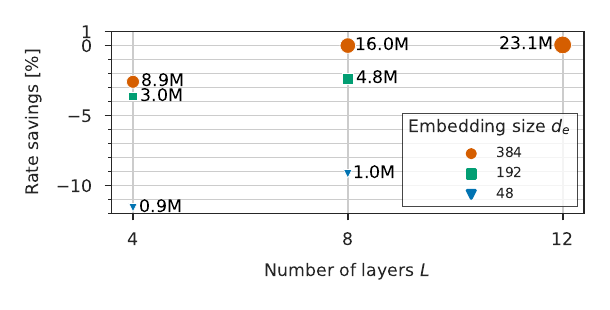}}
	\caption{Various ablation studies conducted with the Contextformer on Kodak dataset. (a) illustrates the rate savings relative to VTM 16.2 for the Contextformer with different number of channel segments $N_{cs}$ and coding order. In (b), the percentile bit occupation per channel segment is shown for models with different coding orders. Each model index depicts a model trained for a specific $\lambda$. Notably, increasing the model capacity allocates more bits in the first two segments for $cfo$ variant. (c) is the illustration of the average BD-Rate performance of various model sizes relative to base model. The annotations indicate the total number of entropy and context model parameters.}
	\label{fig:res345}
\end{figure}

In order to show the generalization capability, we also evaluated  our Contexformer on CLIC2020-Professional and -Mobile \cite{CLIC2020}, and Tecnick \cite{asuni2014testimages} datasets. In terms of BD-Rate \cite{bjontegaard2001calculation}, our model achieves average savings of $9.8\%$, $5.8\%$, and $10.5\%$ over VTM 16.2 on those datasets, respectively (see \cref{fig:res345}). Evaluating on the generalized BD-Rate metric~\cite{minnen2020channel} reveals that our method provides up to $11.9\%$ and $6.6\%$ relative bit savings over VTM~16.2 on CLIC2020-Professional and -Mobile datasets. On Tecnick dataset, the Contextformer saves up to $12.4\%$ more bits over VTM~16.2.

\begin{figure}
	\centering
	\subfloat[\label{fig:res6}]{\includegraphics[width=0.495\linewidth]{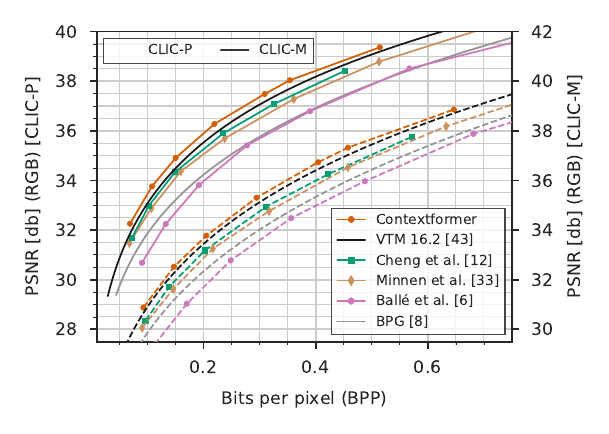}}
	\subfloat[\label{fig:res7}]{\includegraphics[width=0.495\linewidth]{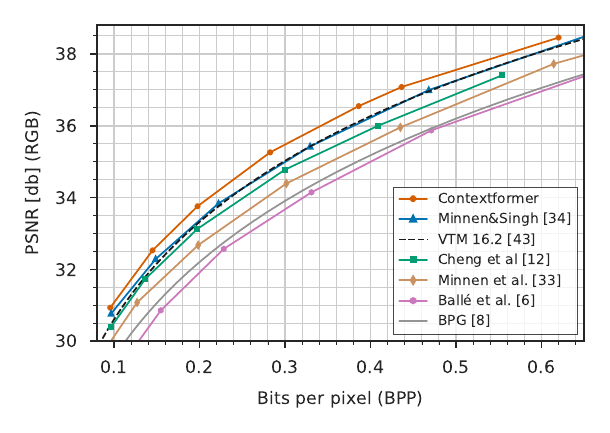}}
	\caption{Comparison of the rate-distortion performance (a) on CLIC2020-Professional (solid line, left vertical axis) and CLIC2020-Mobile (dashed line, right vertical axis) datasets, and (b) on Tecnick dataset.}
	\label{fig:res67}
\end{figure}

\subsubsection{Contextformer Variants.}
In \cref{fig:res3} we show the performance of our model for a varying number of channel segments $N_{cs}$ and coding order. From the figure, one can see that increasing the number of channel segments increases the performance of the model since having more channel segments allows the models to explore more of the cross-channel dependencies. However, there is a trade-off between the number of segments and the complexity -- the computational cost increases quadratically with raising $N_{cs}$. According to our observations, training the Contextformer with more than four segments increases training complexity too much to justify the minor performance gains.

On average, the Contextformer\,$(cfo)$ outperforms the same model in a spatial-first-order ($sfo$) configuration, which highlights the greater importance of cross-channel dependencies than the spatial ones. For instance, in the Contextformer \,$(sfo)$, the primary channel segment can only adopt spatial attention due to the coding order. In \cref{fig:res4} we show the distribution of information in each channel segment. The Contextformer\,$(cfo)$ stores more than $70\%$ of all information in the first two channel segments, while in Contextformer$(sfo)$ the information is almost equally distributed along with the segments. We observed that the spatial-first coding provides a marginal gain in low target bitrates ($bpp\,{<}\,0.3$) and images with a uniformly distributed texture such as ``kodim02'' (in Kodak image dataset). This suggests that spatial dependencies become more pronounced in smoother images. 

\subsubsection{Model Size.}
\cref{fig:res5} shows the performance of the Contextformers for different model sizes compared to the base configuration. Change in the network depth $L$ and embedding size $d_e$ have similar effects on the performance, whereas best performance can be achieved when both are increased. However, we observed that the return diminishes after a network depth of 8 layers. Since the base model already achieves state-of-the-art performance and further upscaling of models increases the complexity, we did not experiment with larger models. Note that the proposed network of~\cite{cui2021asymmetric}, which our model is based on, has approximately the same total number of entropy and context model parameters (17M) as our base model, whereas our model shows additionally $10.1\%$ BD-Rate coding gain on Kodak dataset.

\subsubsection{Runtime Complexity.}
\cref{tab:complexity} shows the encoding and decoding complexity of our model, some of the learning based- prior arts and VTM~\cite{jvet2019versatile}. We tested the learning-based methods on a single NVIDIA Titan RTX, and ran the VTM~\cite{jvet2019versatile} on Intel Core i9-10980XE Intel Core i9-10980XE. In our model, we used proposed BDS and SCS optimizations in the encoder and wavefront coding in the decoder. For low resolution images, our methods have close performance to the one of the 3D context. For 4K images, we observed even bigger benefits by the parallelization. The relative encoding time increases only 3x w.r.t. the one on the Kodak dataset, while the increase in number of pixels is 20-fold. Such speed-up shows that encoder methods with online rate-distortion optimization such as \cite{zhao2021universal} have unexplored potential. Moreover, we achieve 9x faster decoding compared to a 3D context model with the proposed wavefront coding. 
\addtolength{\tabcolsep}{6pt}    
\begin{table}
	\centering
	\caption{Encoding and decoding time of different compression frameworks.}
	\begin{tabular}{@{}lcccc@{}}
		\toprule
		& \multicolumn{2}{c}{Enc. Time [s]}& \multicolumn{2}{c}{Dec. Time [s]} \\
		\cmidrule(lr){2-3} 		\cmidrule(lr){4-5}
		Method & Kodak & 4K&Kodak &4K \\
		\midrule
		DS                                 & \phantom{0}56 	& 1240			 & \phantom{0}62\phantom{.0} &1440\phantom{.0} \\
		BDS (ours)                         & \phantom{0}32 	& \phantom{0}600 & \textendash				 &\textendash \\
		BDS\&SCS (ours)                    & \phantom{00}8  & \phantom{0}120 & \textendash				 &\textendash \\
		Wavefront (ours)                   & \phantom{0}40 	& \phantom{0}760 & \phantom{0}44\phantom{.0} & \phantom{0}820\phantom{.0} \\
		\midrule
		3D context \cite{liu2019non}       & \phantom{00}4	& \phantom{00}28 & 316\phantom{.0}			 &7486\phantom{.0} \\
		2D context \cite{cheng2020learned} & \phantom{00}2	& \phantom{00}54 & \phantom{00}6\phantom{.0} &\phantom{0}140\phantom{.0} \\
		VTM 16.2 \cite{jvet2019versatile}  & 420 			& \phantom{0}950 & \phantom{00}0.8			 &\phantom{000}2.5 \\
		\bottomrule
	\end{tabular}
	\label{tab:complexity}
\end{table}
\addtolength{\tabcolsep}{-6pt}
\section{Conclusion}
In this work, we explored learned image compression architectures using a trans\-former-based context model. We proposed a context model that utilizes a multi-head attention mechanism and uses spatial dependencies in the latent space to model the entropy. Additionally, we also proposed a more generalized attention mechanism, spatio-channel attention, which can constitute a powerful context model. We showed that a compression architecture that employs the spatio-channel attention model achieves state-of-the-art rate-distortion performance.

While using an entropy model with spatio-channel attention brings noticeable gain, it also increases the runtime complexity. We addressed this issue by proposing an algorithm for efficient modeling while keeping the architecture unchanged. Future work will further investigate efficient attention mechanisms (e.g., \cite{pmlr-v119-katharopoulos20a,lee2021fnet,roy2021efficient}) aiming to bridge the gap to a real-time operation.

\newpage
\bibliographystyle{splncs04}
\bibliography{literature}

\begin{thebibliography}{10}
\providecommand{\url}[1]{\texttt{#1}}
\providecommand{\urlprefix}{URL }
\providecommand{\doi}[1]{https://doi.org/#1}

\bibitem{ohm2018versatile}
{Versatile Video Coding}. Standard, Rec. ITU-T H.266 and ISO/IEC 23090-3 (Aug
  2020)

\bibitem{asuni2014testimages}
Asuni, N., Giachetti, A.: Testimages: a large-scale archive for testing visual
  devices and basic image processing algorithms. In: STAG. pp. 63--70 (2014)

\bibitem{balle2020nonlinear}
Ball{\'e}, J., Chou, P.A., Minnen, D., Singh, S., Johnston, N., Agustsson, E.,
  Hwang, S.J., Toderici, G.: Nonlinear transform coding. IEEE Journal of
  Selected Topics in Signal Processing  \textbf{15}(2),  339--353 (2020)

\bibitem{balle2015density}
Ball{\'e}, J., Laparra, V., Simoncelli, E.P.: Density modeling of images using
  a generalized normalization transformation. In: 4th International Conference
  on Learning Representations, ICLR 2016 (2016)

\bibitem{balle2017end}
Ball{\'e}, J., Laparra, V., Simoncelli, E.P.: End-to-end optimized image
  compression. In: 5th International Conference on Learning Representations,
  ICLR 2017 (2017)

\bibitem{balle2018variational}
Ball{\'e}, J., Minnen, D., Singh, S., Hwang, S.J., Johnston, N.: Variational
  image compression with a scale hyperprior. In: International Conference on
  Learning Representations (2018)

\bibitem{begaint2020compressai}
B{\'e}gaint, J., Racap{\'e}, F., Feltman, S., Pushparaja, A.: Compressai: a
  pytorch library and evaluation platform for end-to-end compression research.
  arXiv preprint arXiv:2011.03029  (2020)

\bibitem{bellard2015bpg}
Bellard, F.: Bpg image format (2015), accessed: 2021-11-05. URL
  \url{https://bellard.org/bpg}

\bibitem{bjontegaard2001calculation}
Bjontegaard, G.: Calculation of average psnr differences between rd-curves.
  VCEG-M33  (2001)

\bibitem{carion2020end}
Carion, N., Massa, F., Synnaeve, G., Usunier, N., Kirillov, A., Zagoruyko, S.:
  End-to-end object detection with transformers. In: European Conference on
  Computer Vision. pp. 213--229. Springer (2020)

\bibitem{liu2019non}
Chen, T., Liu, H., Ma, Z., Shen, Q., Cao, X., Wang, Y.: End-to-end learnt image
  compression via non-local attention optimization and improved context
  modeling. IEEE Transactions on Image Processing  \textbf{30},  3179--3191
  (2021). \doi{10.1109/TIP.2021.3058615}

\bibitem{cheng2020learned}
Cheng, Z., Sun, H., Takeuchi, M., Katto, J.: Learned image compression with
  discretized gaussian mixture likelihoods and attention modules. In:
  Proceedings of the IEEE/CVF Conference on Computer Vision and Pattern
  Recognition. pp. 7939--7948 (2020)

\bibitem{cui2020g}
Cui, Z., Wang, J., Bai, B., Guo, T., Feng, Y.: G-vae: A continuously variable
  rate deep image compression framework. arXiv preprint arXiv:2003.02012
  (2020)

\bibitem{cui2021asymmetric}
Cui, Z., Wang, J., Gao, S., Guo, T., Feng, Y., Bai, B.: Asymmetric gained deep
  image compression with continuous rate adaptation. In: Proceedings of the
  IEEE/CVF Conference on Computer Vision and Pattern Recognition. pp.
  10532--10541 (2021)

\bibitem{devlin-etal-2019-bert}
Devlin, J., Chang, M.W., Lee, K., Toutanova, K.: {BERT}: Pre-training of deep
  bidirectional transformers for language understanding. In: Proceedings of the
  2019 Conference of the North {A}merican Chapter of the Association for
  Computational Linguistics: Human Language Technologies, Volume 1 (Long and
  Short Papers). pp. 4171--4186. Association for Computational Linguistics,
  Minneapolis, Minnesota (Jun 2019). \doi{10.18653/v1/N19-1423},
  \url{https://aclanthology.org/N19-1423}

\bibitem{dosovitskiy2020image}
Dosovitskiy, A., Beyer, L., Kolesnikov, A., Weissenborn, D., Zhai, X.,
  Unterthiner, T., Dehghani, M., Minderer, M., Heigold, G., Gelly, S., et~al.:
  An image is worth 16x16 words: Transformers for image recognition at scale.
  In: International Conference on Learning Representations (2020)

\bibitem{esser2021taming}
Esser, P., Rombach, R., Ommer, B.: Taming transformers for high-resolution
  image synthesis. In: Proceedings of the IEEE/CVF Conference on Computer
  Vision and Pattern Recognition. pp. 12873--12883 (2021)

\bibitem{franzen1999kodak}
Franzen, R.: Kodak lossless true color image suite  (1999)

\bibitem{CLIC2020}
George~Toderici, Wenzhe~Shi, R.T.L.T.J.B.E.A.N.J.F.M.: Workshop and challenge
  on learned image compression (clic2020) (2020),
  \url{http://www.compression.cc}

\bibitem{goyal2001theoretical}
Goyal, V.K.: Theoretical foundations of transform coding. IEEE Signal
  Processing Magazine  \textbf{18}(5),  9--21 (2001)

\bibitem{guo2021causal}
Guo, Z., Zhang, Z., Feng, R., Chen, Z.: Causal contextual prediction for
  learned image compression. IEEE Transactions on Circuits and Systems for
  Video Technology  (2021)

\bibitem{he2021checkerboard}
He, D., Zheng, Y., Sun, B., Wang, Y., Qin, H.: Checkerboard context model for
  efficient learned image compression. In: Proceedings of the IEEE/CVF
  Conference on Computer Vision and Pattern Recognition. pp. 14771--14780
  (2021)

\bibitem{jiang2021transgan}
Jiang, Y., Chang, S., Wang, Z.: Transgan: Two transformers can make one strong
  gan. arXiv preprint arXiv:2102.07074  (2021)

\bibitem{pmlr-v119-katharopoulos20a}
Katharopoulos, A., Vyas, A., Pappas, N., Fleuret, F.: Transformers are {RNN}s:
  Fast autoregressive transformers with linear attention. In: III, H.D., Singh,
  A. (eds.) Proceedings of the 37th International Conference on Machine
  Learning. Proceedings of Machine Learning Research, vol.~119, pp. 5156--5165.
  PMLR (13--18 Jul 2020),
  \url{https://proceedings.mlr.press/v119/katharopoulos20a.html}

\bibitem{kim2019grdn}
Kim, D.W., Ryun~Chung, J., Jung, S.W.: Grdn: Grouped residual dense network for
  real image denoising and gan-based real-world noise modeling. In: Proceedings
  of the IEEE/CVF Conference on Computer Vision and Pattern Recognition
  Workshops. pp.~0--0 (2019)

\bibitem{kingma2014adam}
Kingma, D.P., Ba, J.: Adam: A method for stochastic optimization. arXiv
  preprint arXiv:1412.6980  (2014)

\bibitem{burak2021}
Koyuncu, A.B., Cui, K., Boev, A., Steinbach, E.: Parallelized context modeling
  for faster image coding. In: 2021 International Conference on Visual
  Communications and Image Processing (VCIP). pp.~1--5. IEEE (2021)

\bibitem{lee2018context}
Lee, J., Cho, S., Beack, S.K.: Context-adaptive entropy model for end-to-end
  optimized image compression. In: 6th International Conference on Learning
  Representations, ICLR 2018 (2018)

\bibitem{lee2021fnet}
Lee-Thorp, J., Ainslie, J., Eckstein, I., Ontanon, S.: Fnet: Mixing tokens with
  fourier transforms. arXiv preprint arXiv:2105.03824  (2021)

\bibitem{li2021involution}
Li, D., Hu, J., Wang, C., Li, X., She, Q., Zhu, L., Zhang, T., Chen, Q.:
  Involution: Inverting the inherence of convolution for visual recognition.
  In: Proceedings of the IEEE/CVF Conference on Computer Vision and Pattern
  Recognition. pp. 12321--12330 (2021)

\bibitem{9067005}
Li, M., Ma, K., You, J., Zhang, D., Zuo, W.: Efficient and effective
  context-based convolutional entropy modeling for image compression. IEEE
  Transactions on Image Processing  \textbf{29},  5900--5911 (2020).
  \doi{10.1109/TIP.2020.2985225}

\bibitem{liu2019practical}
Liu, H., Chen, T., Shen, Q., Ma, Z.: Practical stacked non-local attention
  modules for image compression. In: CVPR Workshops. p.~0 (2019)

\bibitem{mentzer2018conditional}
Mentzer, F., Agustsson, E., Tschannen, M., Timofte, R., Van~Gool, L.:
  Conditional probability models for deep image compression. In: Proceedings of
  the IEEE Conference on Computer Vision and Pattern Recognition. pp.
  4394--4402 (2018)

\bibitem{minnen2018joint}
Minnen, D., Ball{\'e}, J., Toderici, G.: Joint autoregressive and hierarchical
  priors for learned image compression. In: NeurIPS (2018)

\bibitem{minnen2020channel}
Minnen, D., Singh, S.: Channel-wise autoregressive entropy models for learned
  image compression. In: 2020 IEEE International Conference on Image Processing
  (ICIP). pp. 3339--3343. IEEE (2020)

\bibitem{naseer2021intriguing}
Naseer, M., Ranasinghe, K., Khan, S., Hayat, M., Khan, F.S., Yang, M.H.:
  Intriguing properties of vision transformers. arXiv preprint arXiv:2105.10497
   (2021)

\bibitem{NIU202148}
Niu, Z., Zhong, G., Yu, H.: A review on the attention mechanism of deep
  learning. Neurocomputing  \textbf{452},  48--62 (2021).
  \doi{https://doi.org/10.1016/j.neucom.2021.03.091},
  \url{https://www.sciencedirect.com/science/article/pii/S092523122100477X}

\bibitem{parmar2018image}
Parmar, N., Vaswani, A., Uszkoreit, J., Kaiser, L., Shazeer, N., Ku, A., Tran,
  D.: Image transformer. In: International Conference on Machine Learning. pp.
  4055--4064. PMLR (2018)

\bibitem{qian2021entroformer}
Qian, Y., Sun, X., Lin, M., Tan, Z., Jin, R.: Entroformer: A transformer-based
  entropy model for learned image compression. In: International Conference on
  Learning Representations (2021)

\bibitem{qian2020learning}
Qian, Y., Tan, Z., Sun, X., Lin, M., Li, D., Sun, Z., Hao, L., Jin, R.:
  Learning accurate entropy model with global reference for image compression.
  In: International Conference on Learning Representations (2020)

\bibitem{rissanen1979arithmetic}
Rissanen, J., Langdon, G.G.: Arithmetic coding. IBM Journal of research and
  development  \textbf{23}(2),  149--162 (1979)

\bibitem{roy2021efficient}
Roy, A., Saffar, M., Vaswani, A., Grangier, D.: Efficient content-based sparse
  attention with routing transformers. Transactions of the Association for
  Computational Linguistics  \textbf{9},  53--68 (2021)

\bibitem{articleewr}
Sullivan, G., Ohm, J., Han, W.J., Wiegand, T.: Overview of the high efficiency
  video coding standard  \textbf{22},  1648--1667 (12 2012)

\bibitem{jvet2019versatile}
Team, J.V.E.: Versatile video coding (vvc) reference software: Vvc test model
  (vtm) (2021), accessed: 2021-11-05. URL
  \url{https://vcgit.hhi.fraunhofer.de/jvet/VVCSoftware\_VTM}

\bibitem{vaswani2017attention}
Vaswani, A., Shazeer, N., Parmar, N., Uszkoreit, J., Jones, L., Gomez, A.N.,
  Kaiser, {\L}., Polosukhin, I.: Attention is all you need. In: Advances in
  neural information processing systems. pp. 5998--6008 (2017)

\bibitem{wallace1992jpeg}
Wallace, G.K.: The jpeg still picture compression standard. IEEE transactions
  on consumer electronics  \textbf{38}(1),  xviii--xxxiv (1992)

\bibitem{wang2003scale}
Wang, Z., Simoncelli, E.P., Bovik, A.C.: Multiscale structural similarity for
  image quality assessment. In: The Thrity-Seventh Asilomar Conference on
  Signals, Systems \& Computers, 2003. vol.~2, pp. 1398--1402. Ieee (2003)

\bibitem{wang2003multiscale}
Wang, Z., Simoncelli, E.P., Bovik, A.C.: Multiscale structural similarity for
  image quality assessment. In: The Thrity-Seventh Asilomar Conference on
  Signals, Systems \& Computers, 2003. vol.~2, pp. 1398--1402. Ieee (2003)

\bibitem{xue2019video}
Xue, T., Chen, B., Wu, J., Wei, D., Freeman, W.T.: Video enhancement with
  task-oriented flow. International Journal of Computer Vision (IJCV)
  \textbf{127}(8),  1106--1125 (2019)

\bibitem{zhang2018residual}
Zhang, Y., Li, K., Li, K., Zhong, B., Fu, Y.: Residual non-local attention
  networks for image restoration. In: International Conference on Learning
  Representations (2019), \url{https://openreview.net/forum?id=HkeGhoA5FX}

\bibitem{zhao2021universal}
Zhao, J., Li, B., Li, J., Xiong, R., Lu, Y.: A universal encoder rate
  distortion optimization framework for learned compression. In: Proceedings of
  the IEEE/CVF Conference on Computer Vision and Pattern Recognition. pp.
  1880--1884 (2021)

\bibitem{zhou2019multi}
Zhou, J., Wen, S., Nakagawa, A., Kazui, K., Tan, Z.: Multi-scale and
  context-adaptive entropy model for image compression. arXiv preprint
  arXiv:1910.07844  (2019)

\bibitem{ziv1985universal}
Ziv, J.: On universal quantization. IEEE Transactions on Information Theory
  \textbf{31}(3),  344--347 (1985)

\end{thebibliography}
\newpage

\setcounter{figure}{0}  
\setcounter{equation}{0}
\setcounter{table}{0}  
\setcounter{section}{0}    
\setcounter{subsection}{0}  
\renewcommand{\thefigure}{A\arabic{figure}}
\renewcommand{\thetable}{A\arabic{table}}
\renewcommand{\thesection}{A\arabic{section}}
\renewcommand{\thealgocf}{A\arabic{algocf}}

\begin{center}%
	\let\newline\\
	{\Large \bfseries\boldmath
		\pretolerance=10000
		Appendix \par}\vskip .8cm
\end{center}%

\section{Illustration of the Runtime Optimization Methods}
\begin{figure}
	\centering
	\subfloat[\label{fig:ours1}Step n]{		\includegraphics[width=0.32\textwidth]{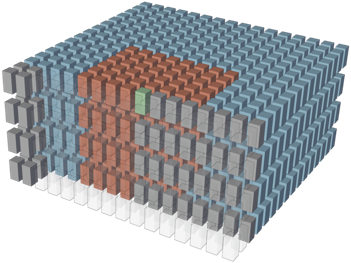}}
	\subfloat[\label{fig:ours2}Step n+3]{		\includegraphics[width=0.32\textwidth]{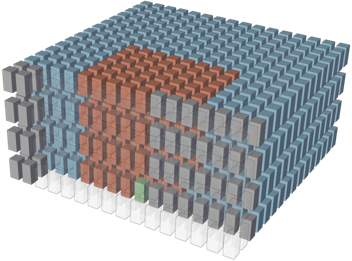}}
	\subfloat[\label{fig:ours3}Step n+7]{		\includegraphics[width=0.32\textwidth]{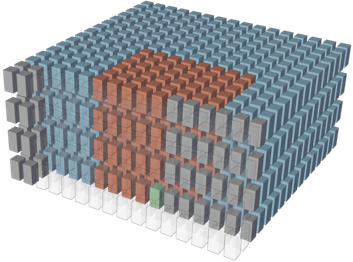}}
	\caption[]{Illustration of the context modeling in a Contextformer$(cfo)$ with channel-first-order processing and sliding window attention for various processing steps $n$. The latent tensors are displayed in different colors:
		{\drawpatternColor{mygreen}} the current latent elements to be coded; {\drawpatternColor{myorange}} the latent elements used by the context model; {\drawpatternColor{myblue}} previously coded elements; and {\drawpatternColor{mygrey}} elements yet to be coded.	  
	}
	\label{fig:all_ctx2}
\end{figure}

\begin{figure}
	\centering
	\includegraphics[width=\linewidth]{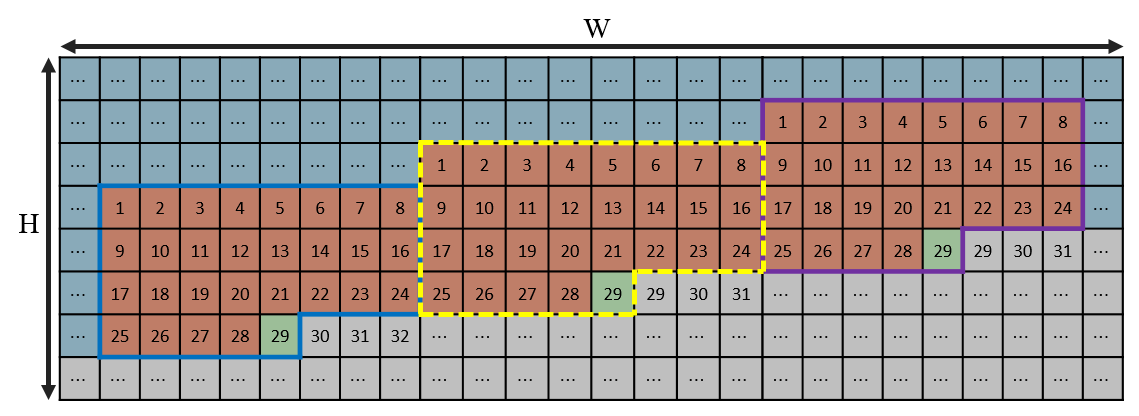}
	\caption[]{Illustration of the context modeling in a Contextformer with sliding window attention and wavefront processing for an arbitrary processing step $n$. The processing order of each latent element is $n+N$, where $N$ is the number given in the figure for each latent. Simultaneously processed windows for step $n+29$ are framed in different colors. For simplicity, the channel dimension is omitted.
	}
	\label{fig:all_ctx3}
\end{figure}

This section gives further details about the proposed runtime optimization methods; \textit{skip intermediate channel segments} (SCS) and \textit{batched dynamic sequence processing} (BDS). \cref{fig:all_ctx2} illustrates a few of the processing steps of a Context\-former\,$(cfo)$. Since the transformers are sequence-to-sequence models, processing step $n+3$ also calculates the output for step $n$. The provided mask allows using of subsequent elements for calculating the attention for each element. Thus, the calculation of the output for step $n$ can be skipped during the encoding without harming the causality (SCS). In the figure, one can see that steps $n+3$ and $n+7$ contain equal number of latent elements used for context modeling, which enables processing them in batches efficiently (BDS). Our algorithm searches for latent elements, which can be grouped into a batch, and then models the context accordingly (see \cref{algo:algo1}).
\cref{fig:all_ctx3} illustrates the wavefront processing for the decoding runtime optimization. Since latent tensor elements with the same processing order can be encoded and decoded independently, we process them in batches. The only requirement for using wavefront processing in the decoder is that the latent tensor elements are coded into the bitstream in the same order as in the encoder. That way, the entropy parameters can be computed regardless of the algorithm optimizing the encoding and ordered according to the wavefront processing index prior to bitstream coding.

\begin{algorithm}
	\caption{Runtime optimization for Contextformer}
	\label{algo:algo1}
	\DontPrintSemicolon
	\SetAlgoLined
	\SetNoFillComment
	\SetNoFillComment
	\KwIn{Seq. Latent Tensor $\bm{\hat{y}_s}$, Contextformer $CTX$}
	\KwOut{Contextformer Output $\bm{\psi}$}
	$Q_{ctx},Q_{crt} \leftarrow \{\}$  \tcp{initialize empty dictionaries}
	$\bm{\psi} \leftarrow \bm{0}$ \tcp{initialize Context Model Output}
	$H,W,N_{cs} \leftarrow get\_original\_shape(\bm{\hat{y}_s})$\;
	\tcp{\hspace{-0.5em}loop over sliding windows \& store indices}
	\ForEach{coord. (i,j,k) \textbf{in} 3DMeshCube($H$,$W$,$N_{cs}$)}{
		\tcp{\hspace{-0.5em}get indices residing in current window}
		$indices \leftarrow get\_indices(i, j, k)$\;
		\tcp{\hspace{-0.5em}compute priority, e.g. $length(idx)$ for BDS}
		$priority \leftarrow get\_priority(i,j,k)$\;
		\tcp{\hspace{-0.5em}store current latent index}
		$Q_{crt}[priority].append((i, j, k))$\; 
		\tcp{\hspace{-0.5em}store latent indices used as context}
		$Q_{ctx}[priority].append(indices)$\;
	}
	\tcp{\hspace{-0.5em}loop over coding priorities \& run CTX}
	\ForEach{priority \textbf{in} sorted($Q_{crr}$.keys())}{
		\tcp{\hspace{-0.5em}restore latent indices}
		$i_{crt} \leftarrow  Q_{crt}[priority]$\;
		$i_{ctx} \leftarrow  Q_{ctx}[priority]$\;
		\tcp{\hspace{-0.5em}run context model for the current latents}
		$\bm{\hat{y}}_{ctx}  \leftarrow  \bm{\hat{y}}_s[i_{ctx}]$\;
		$\bm{\psi}[i_{crt}]  \leftarrow  CTX(\bm{\hat{y}}_{ctx} )$\;
	}
\end{algorithm}

\section{Architectural Details}
\cref{tab:model_description} outlines the architectural details of the compression model with the Contextformer.
\setlength{\tabcolsep}{4pt}
\begin{table}[!htbp] 
	\centering
	\caption{An overview of the proposed model with the Contextformer, where each row depicts one layer or component of the model. ``Conv: $K \,{\times}\, K \,{\times}\, N\,\text{s}2$'' stands for a convolutional layer with kernel size of $K \,{\times}\, K$, number of $N$ output channels, and stride ``s'' of 2. Similarly, ``Deconv'' is an abbreviation for transposed convolutions. IGDN is the inverse function of GDN~\cite{balle2015density}. ``Dense'' layers are specified by their output dimension, whereas $K_1\,{=}\,2M \,{+}\, d_e$, $K_2\,{=}\,(4k_{m}M)/N_{cs}$. We selected the base configuration of the Contextformer as $\{L\,{=}\,8,\,d_e\,{=}\,384,\,d_{mlp}\,{=}\,4d_e,\,h\,{=}\,12,\,N_{cs}\,{=}\,4,\,co \,{=}\, cfo\}$	
	}
	\scriptsize
	\begin{tabular}{@{}cccc@{}}
		\toprule[0.8pt]
		Encoder & Decoder & Hyperencoder & Hyperdecoder \\
		\midrule[0.4pt]
		Conv: $3{\times}3{\times}N\;{\text{s}}2$ & $2{\times}$ResBlock: $3{\times}3{\times}M$ & Conv: $3{\times}3{\times}N\;{\text{s}}1$ &  Conv: $3{\times}3{\times}M \;{\text{s}}1$ \\
		
		GDN & Deconv: $3{\times}3{\times}N\; {\text{s}2}$ & Leaky ReLU & Deconv: $3{\times}3{\times}M \;{{\text{s}}2}$ \\
		
		RNAB & IGDN & Conv: $3{\times}3{\times}N \;{\text{s}}1$& Leaky ReLU \\
		
		Conv: $3{\times}3{\times}N\;{\text{s}}2$ & Deconv: $3{\times}3{\times}N\;{{\text{s}}2}$ & Conv: $3{\times}3{\times}N\;{\text{s}}2$ & Conv: $3{\times}3{\times}M \;{\text{s}}1$ \\
		
		GDN & IGDN & Leaky ReLU& Deconv: $3{\times}3{\times}\frac{3}{2}M \;{{\text{s}}2}$\\
		
		Conv: $3{\times}3{\times}N\;{\text{s}}2$ & Deconv: $3{\times}3{\times}N\;{{\text{s}}2}$ & Conv: $3{\times}3{\times}N\;{\text{s}}1$ & Leaky ReLU \\
		
		GDN & RNAB & Conv: $3{\times}3{\times}N\;{\text{s}}2$ &Deconv: $3{\times}3{\times}2M \;{{\text{s}}2}$\\
		
		Conv: $3{\times}3{\times}N\;{\text{s}}2$ & IGDN & & Leaky ReLU\\
		
		Conv: $1{\times}1{\times}M\;{\text{s}}2$ & Deconv: $3{\times}3{\times}3\;{{\text{s}}2}$ & & \\
		\bottomrule[0.8pt]
		&&&\\
		\cmidrule[0.8pt]{2-3}
		& Context Model & Entropy Parameters&\\
		\cmidrule[0.4pt]{2-3}
		& Contextformer: & Dense: $(2K_1{+}K_2)/3$& \\
		&$\{L,d_e,\mathit{d}_{mlp},h,N_{cs},co\}$& GELU&\\
		&& Dense: $(K_1{+}2K_2)/3$&\\
		&& GELU&\\
		&& Dense: $K_2$ &\\
		\cmidrule[0.8pt]{2-3}
	\end{tabular}
	\label{tab:model_description}
\end{table}
\setlength{\tabcolsep}{1.4pt}

\section{Details of Experiments}
\subsection{Obtaining Prior-Art Results}
As explained in \cref{sec:ex}, we compare the performance of our model with various compression methods from the prior art. For this purpose, we used the official open-source implementations of those methods for the inference. For all methods, we used their default configurations. The links to those implementations are available in \cref{tab:details}. Where the source is not available, we extracted the results from the corresponding publications. We also evaluated our extraction method by using it for the methods with available sources. We observed a negligible difference ($<0.1\%$ in BD-Rate) between the inferred and extracted results.
\setlength{\tabcolsep}{4pt}
\begin{table}
	\centering
	\caption{Prior-art methods mentioned in this work with their official sources.}
	\footnotesize
	\begin{tabular}{@{}p{0.25\linewidth}p{0.68\linewidth}@{}}
		\toprule
		Method & Link \\
		\midrule
		Minnen et al.~\cite{minnen2018joint} & \multirow{2}{30em}{\url{https://github.com/tensorflow/compression}}\\
		Minnen\&Singh~\cite{minnen2020channel} & \\
		\midrule
		Cheng et al.~\cite{cheng2020learned} & \url{https://github.com/ZhengxueCheng/Learned-Image-Compression-with-GMM-and-Attention}\\
		\midrule	
		Chen et al.~\cite{liu2019non} & \url{		https://github.com/NJUVISION/NIC}\\
		\midrule	
		BPG~\cite{bellard2015bpg} & \url{https://bellard.org/bpg/}\\
		\midrule
		VTM~16.2~\cite{jvet2019versatile}&\url{https://vcgit.hhi.fraunhofer.de/jvet/VVCSoftware_VTM}\\
		\bottomrule
	\end{tabular}
	\label{tab:details}
\end{table}
\setlength{\tabcolsep}{1.4pt}

\subsection{Detailed Performance Comparison}
\begin{figure}
	\centering
	{\large \bf Results on Kodak image dataset (PSNR)}\par\medskip
	\includegraphics[width=1\linewidth]{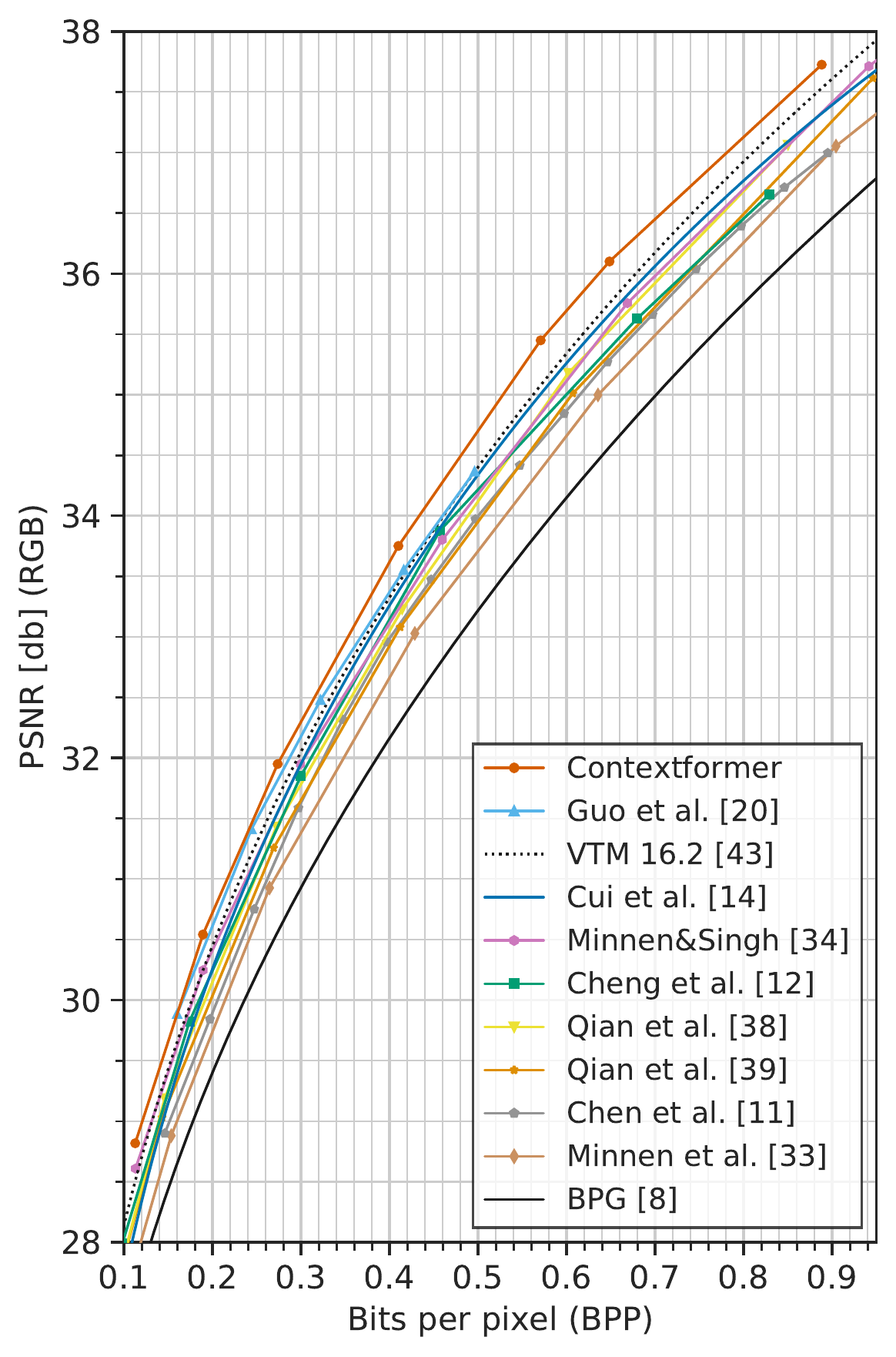}
	
	\caption{\the\linewidth Rate-distortion performance comparison on Kodak dataset in terms of PSNR for our model, and various learning-based and classical codecs.}
	\label{fig:res1_big}
\end{figure}

\begin{figure}
	\centering
	{\large \bf Results on Kodak image dataset (MS-SSIM)}\par\medskip
	\includegraphics[width=1\linewidth]{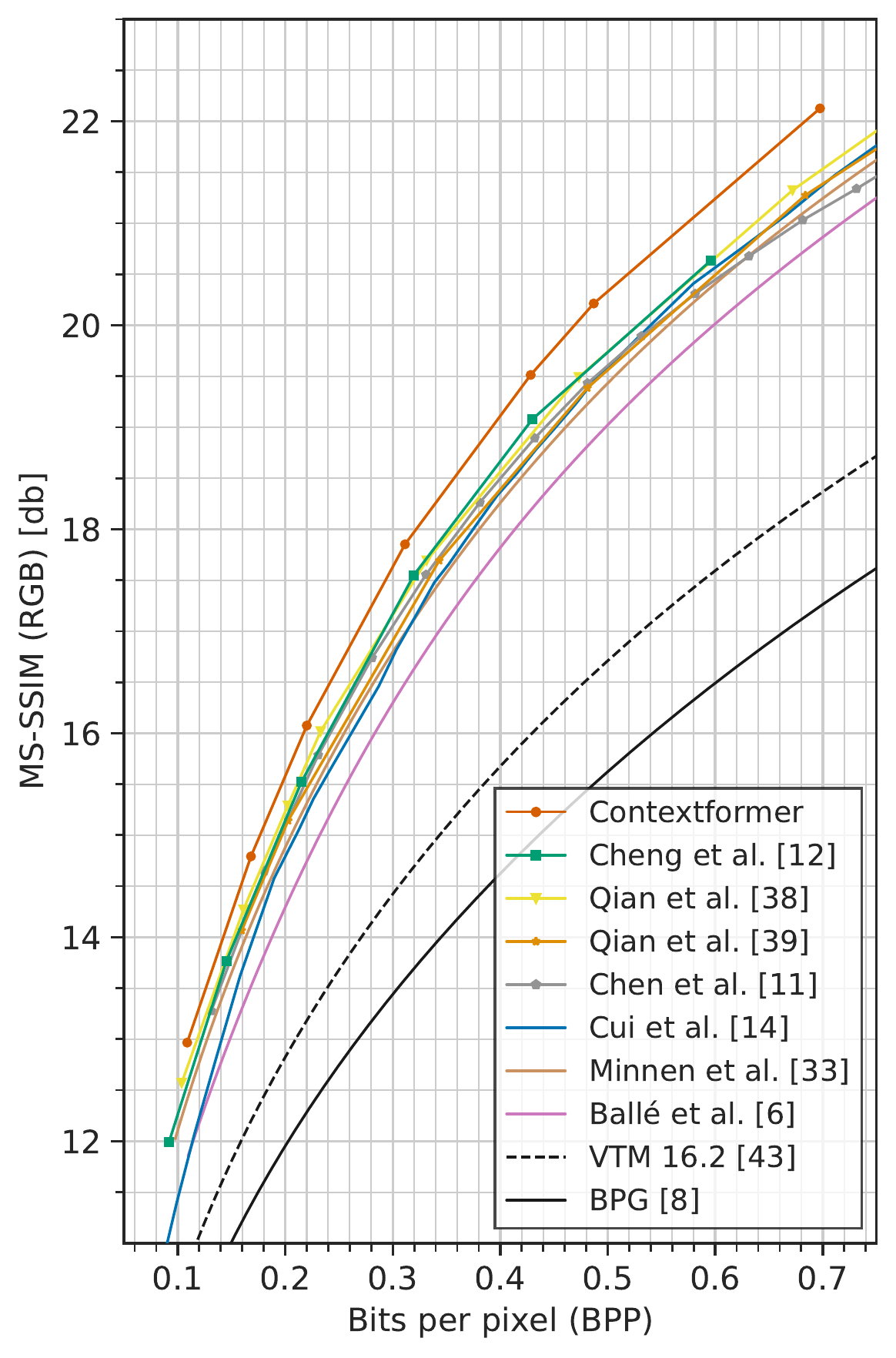}
	
	\caption{Rate-distortion performance comparison on Kodak dataset in terms of MS-SSIM for our model, and various learning-based and classical codecs.}
	\label{fig:res4_big}
\end{figure}

\begin{figure}
	\centering
	{\large \bf Results on CLIC2020 dataset (PSNR)}\par\medskip
	\includegraphics[width=1\linewidth]{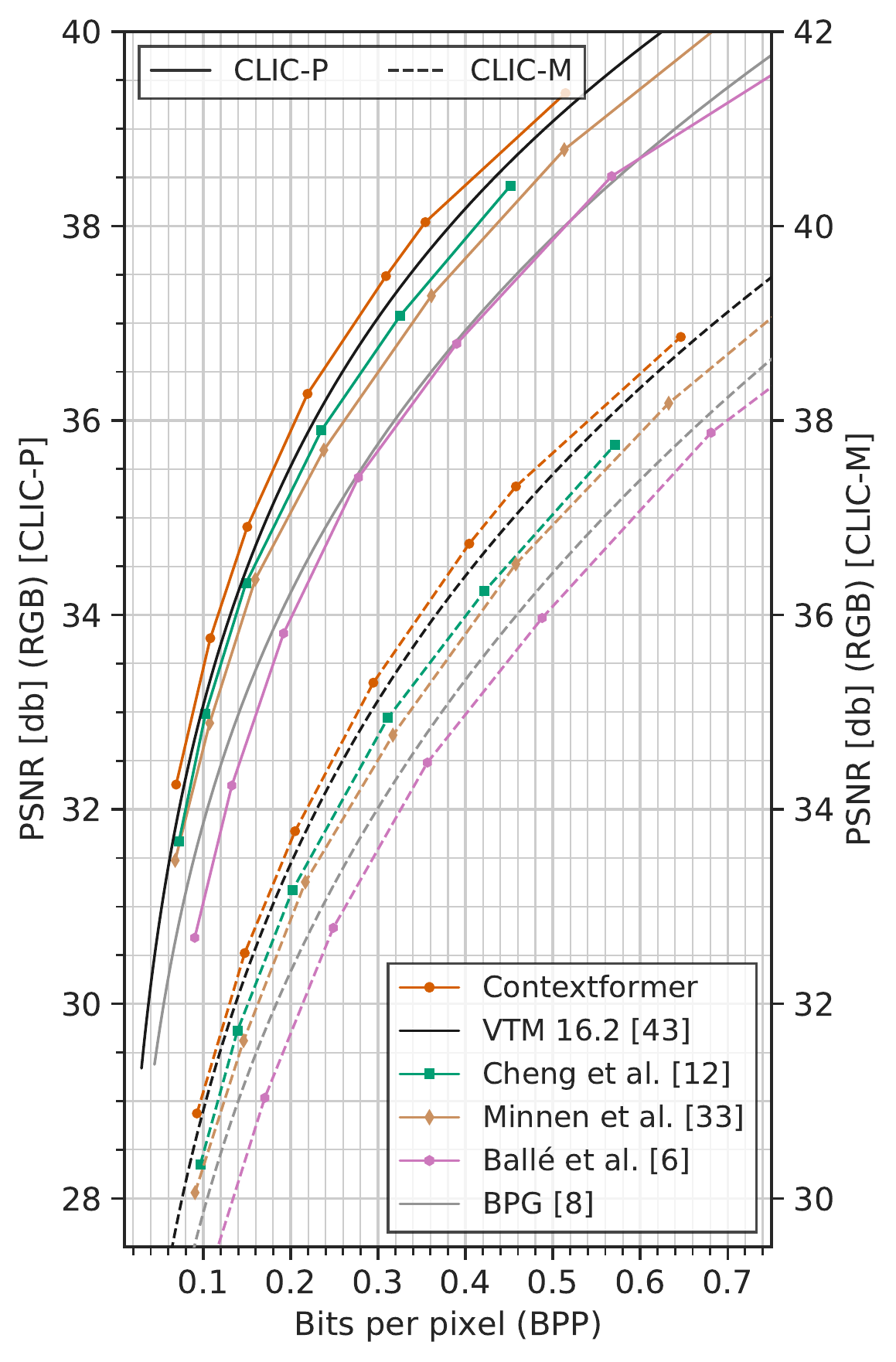}
	
	\caption{Rate-distortion performance comparison on CLIC-Professional (solid line, left vertical axis) and CLIC-Mobile (dashed line, right vertical axis) datasets in terms of PSNR for our model, and various learning-based and classical codecs.}
	\label{fig:res2_big}
\end{figure}

\begin{figure}
	\centering
	{\large \bf Results on CLIC2020 dataset (MS-SSIM)}\par\medskip
	\includegraphics[width=1\linewidth]{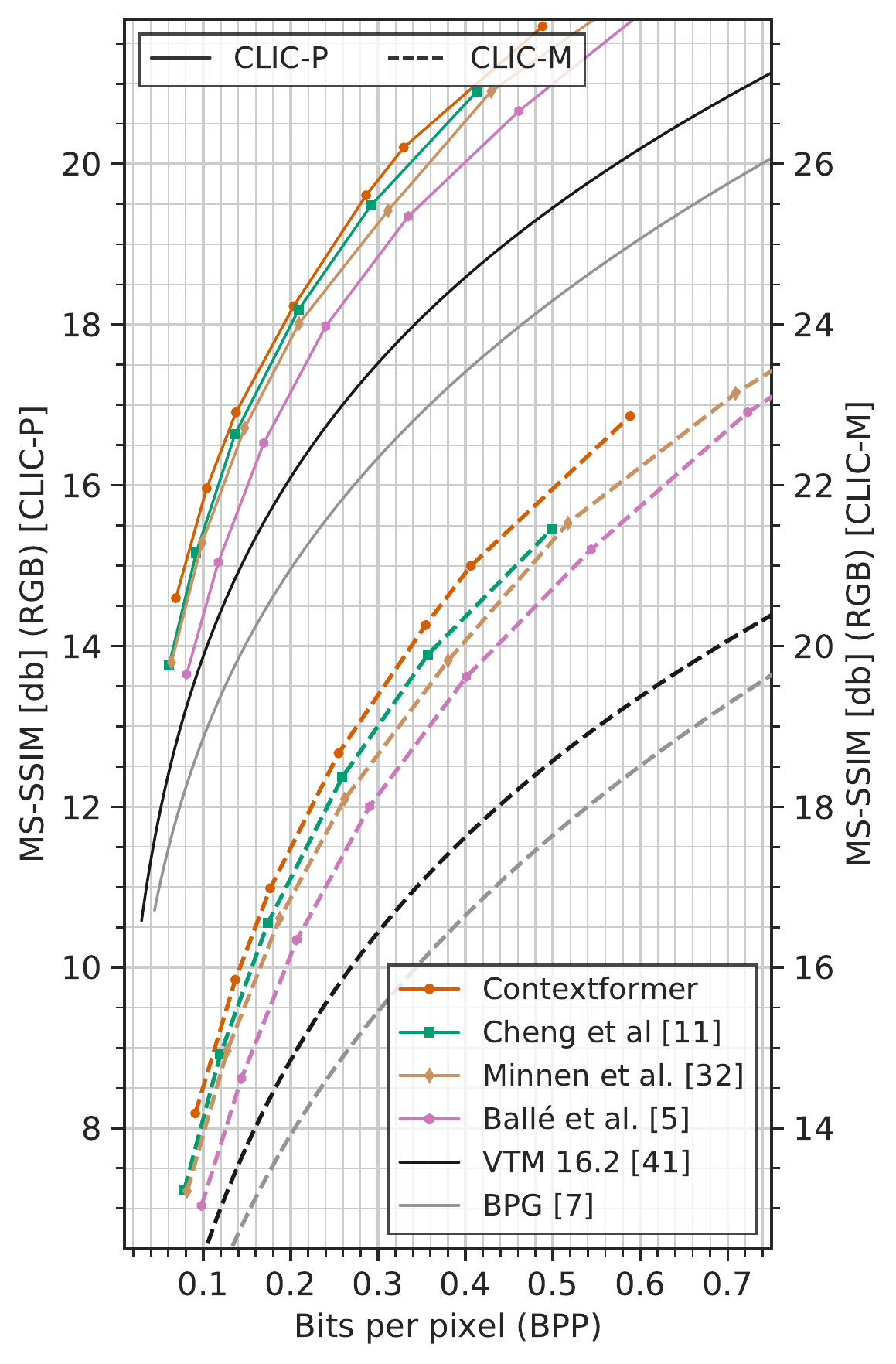}
	
	\caption{Rate-distortion performance comparison on CLIC-Professional (solid line, left vertical axis) and CLIC-Mobile (dashed line, right vertical axis) datasets in terms of MS-SSIM for our model, and various learning-based and classical codecs.}
	\label{fig:res5_big}
\end{figure}

\begin{figure}
	\centering
	{\large \bf Results on Tecnick dataset (PSNR)}\par\medskip
	\includegraphics[width=1\linewidth]{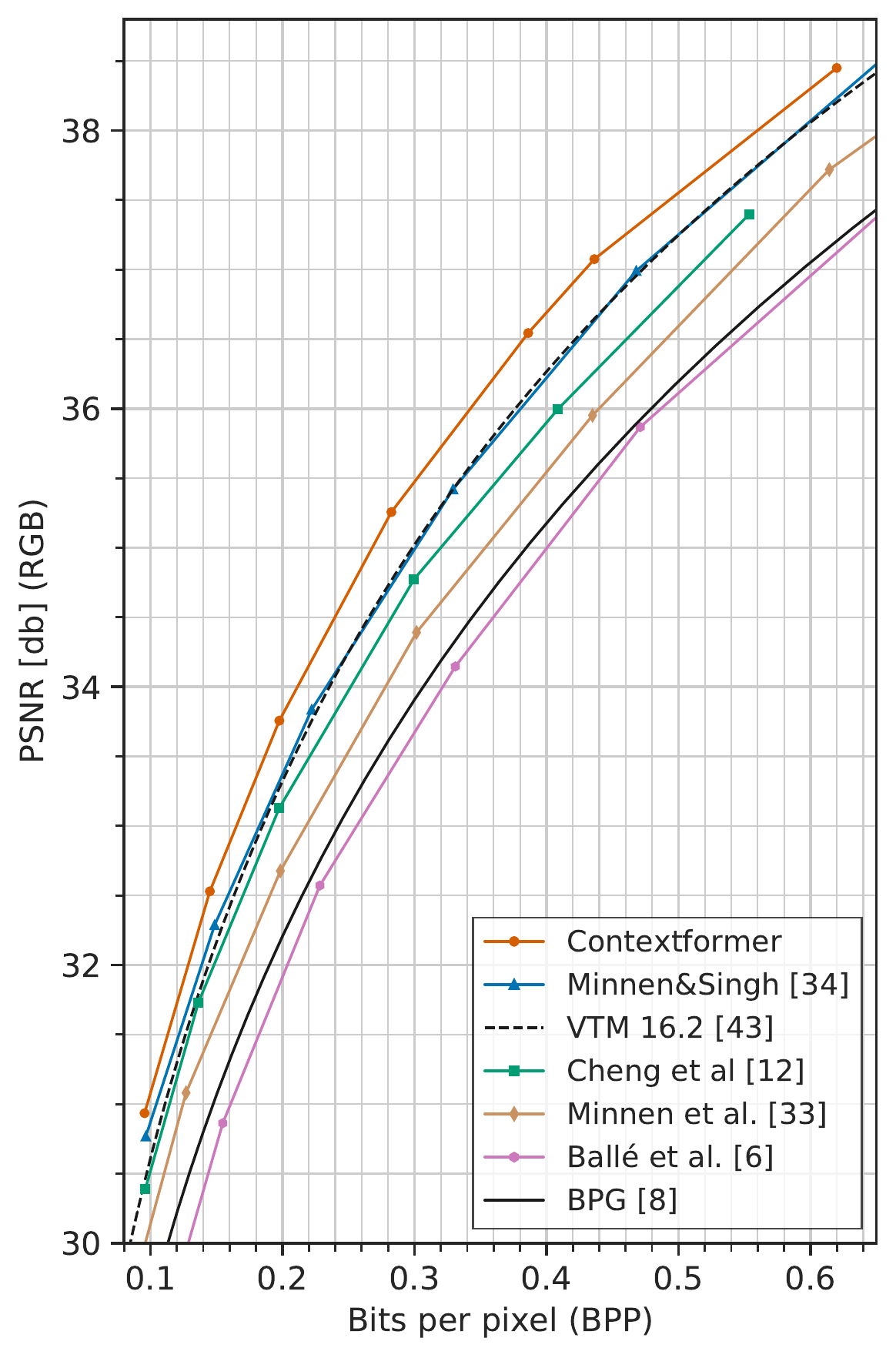}
	
	\caption{Rate-distortion performance comparison on Tecnick dataset in terms of PSNR for our model, and various learning-based and classical codecs.}
	\label{fig:res3_big}
\end{figure}

\begin{figure}
	\centering
	{\large \bf Results on Tecnick dataset (MS-SSIM)}\par\medskip
	\includegraphics[width=1\linewidth]{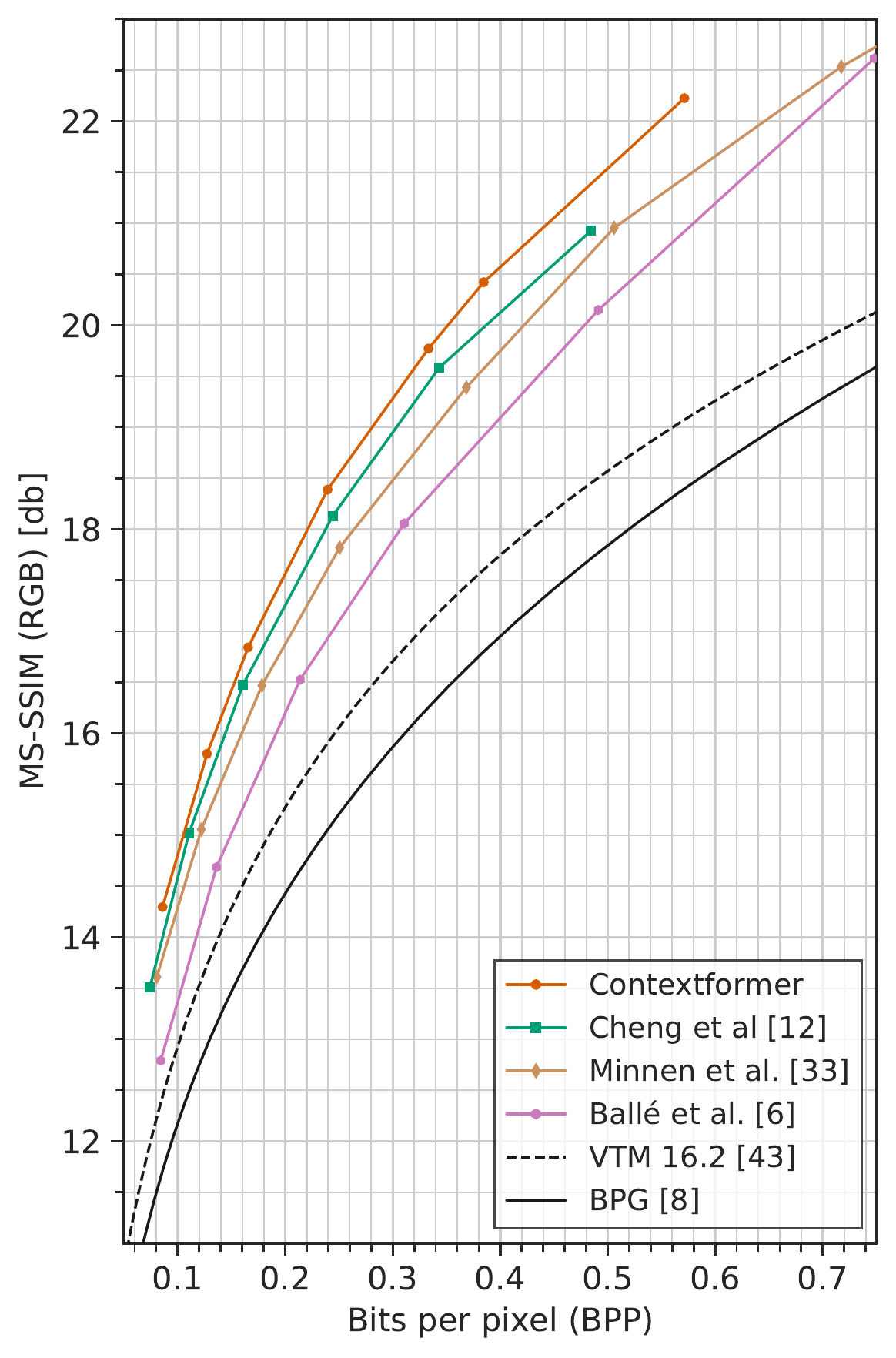}
	
	\caption{Rate-distortion performance comparison on Tecnick dataset in terms of MS-SSIM for our model, and various learning-based and classical codecs.}
	\label{fig:res6_big}
\end{figure}

We provide more extensive versions of the figures presented in~\cref{sec:performance}. \Cref{fig:res1_big,fig:res4_big} show the same performance comparison on Kodak dataset as \cref{fig:res1,fig:res8}, but it includes additional prior-art methods such as Qian et al.~\cite{qian2020learning} and Chen et al.~\cite{liu2019non}. Similarly, \cref{fig:res2_big,fig:res3_big} illustrate the performance comparison on CLIC2020 and Tecnick datasets, which show the same results as \cref{fig:res6,fig:res7} but provide better visuals. For the sake of completeness, we also provide the results on CLIC2020 and Tecnick datasets for our models optimized for MS-SSIM (see \cref{fig:res5_big,fig:res6_big}).

\subsection{Detailed Model Size Comparison}
\cref{tab:modelsize} shows the number of parameters of our compression framework compared to various frameworks. Depending on the implementation, we used the summary functions of PyTorch or Tensorflow for the calculation. Compared to the other transformer-based model such as Qian et al.~\cite{qian2021entroformer}, our model employs relatively complex encoder/decoder with a smaller bottleneck, and much simpler hyperprior. Additionally, our context model with spatio-channel attention requires an order of magnitude smaller parameters compared to the other context model adopting channelwise processing such as Minnen\&Singh~\cite{minnen2020channel}. The total number of parameters in our entropy modeling is relatively smaller than the various approaches including our baseline Cui et al.~\cite{cui2021asymmetric}. Therefore, our method might have a higher potential for adopting online rate-distortion optimization techniques such as \cite{zhao2021universal}.

\addtolength{\tabcolsep}{4pt}    
\begin{table}
	\small
	\centering
	\caption{Number of parameters of various models, which are calculated by summary functions of the used framework. }
	\label{tab:modelsize}
	\begin{tabular}{@{}lccccc@{}}
		\toprule
		&&& Hyper&Hyper& Context+Entropy \\
		Method& Encoder&Decoder & Encoder & Decoder & Parameters \\
		\midrule
		Contextformer & 8.1M & 9.4M &1.6M                         &\phantom{1}2.4M    &\phantom{1}15.9M \\
		Cui et al.~\cite{cui2021asymmetric} & 8.1M & 9.4M &1.6M   &\phantom{1}2.4M    &\phantom{1}17.2M \\
		Qian et al.~\cite{qian2021entroformer} & 3.8M & 3.8M &8.2M&15.9M      &\phantom{1}13.1M \\
		Minnen\&Singh \cite{minnen2020channel} & 2.8M & 2.8M &2.1M&\phantom{1}2.1M &\phantom{10}2.8M \\	
		Minnen et al.~\cite{minnen2018joint} & 4.2M & 4.2M &5.2M  &\phantom{1}5.8M   &101.9M \\
		\bottomrule
	\end{tabular}
\end{table}
\addtolength{\tabcolsep}{-4pt}

\subsection{Visual Assessment}
\begin{figure}
	\centering
	\includegraphics[width=1\linewidth]{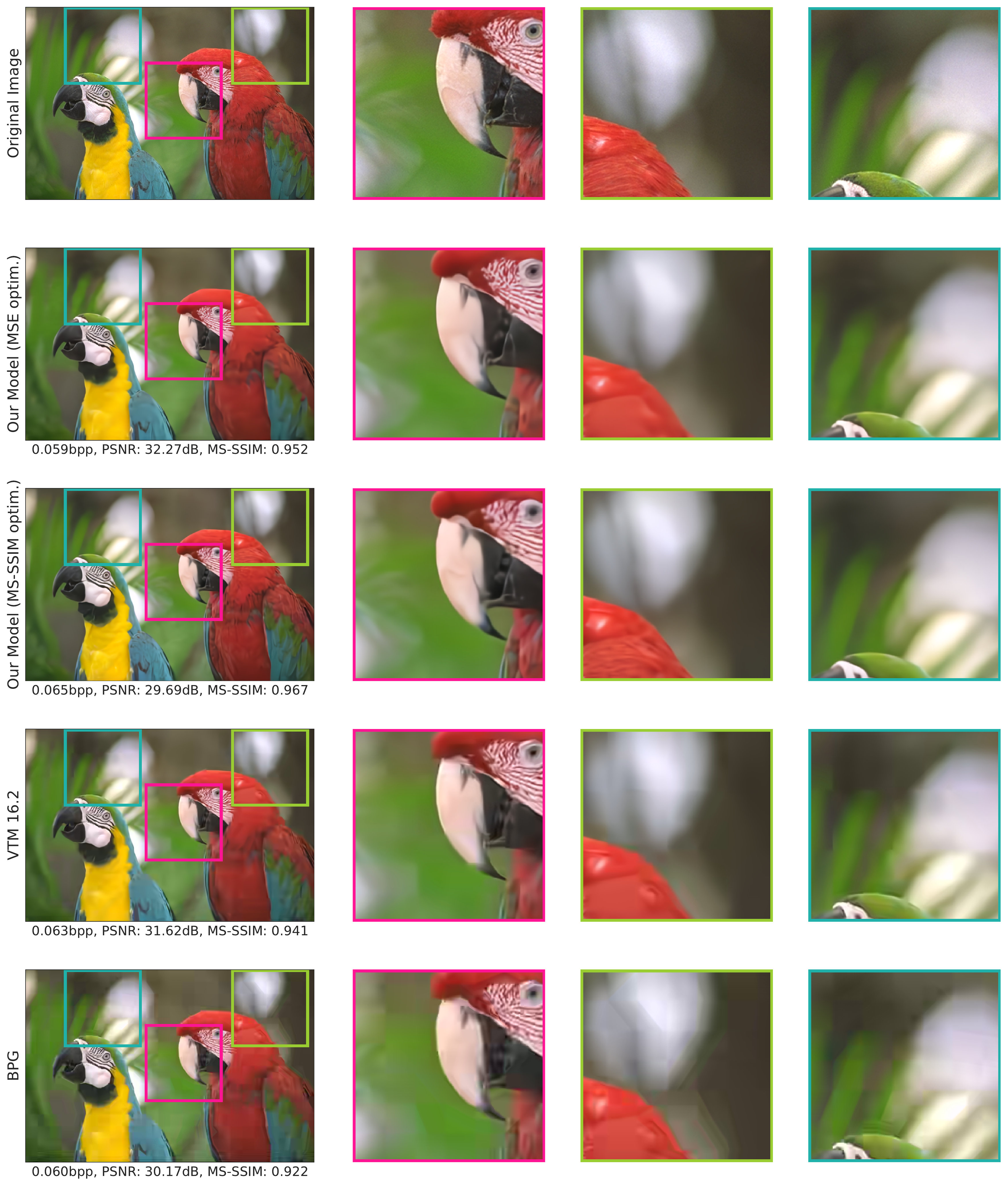}
	
	\caption{The reconstructed images of \textit{kodim23} from the Kodak image dataset for the visual comparison. The image is compressed by our models (optimized for MSE or MS-SSIM), VTM 16.2, and BPG for the target bpp of 0.06.}
	\label{fig:viz2}
\end{figure}

\begin{figure}
	\centering
	\includegraphics[width=1\linewidth]{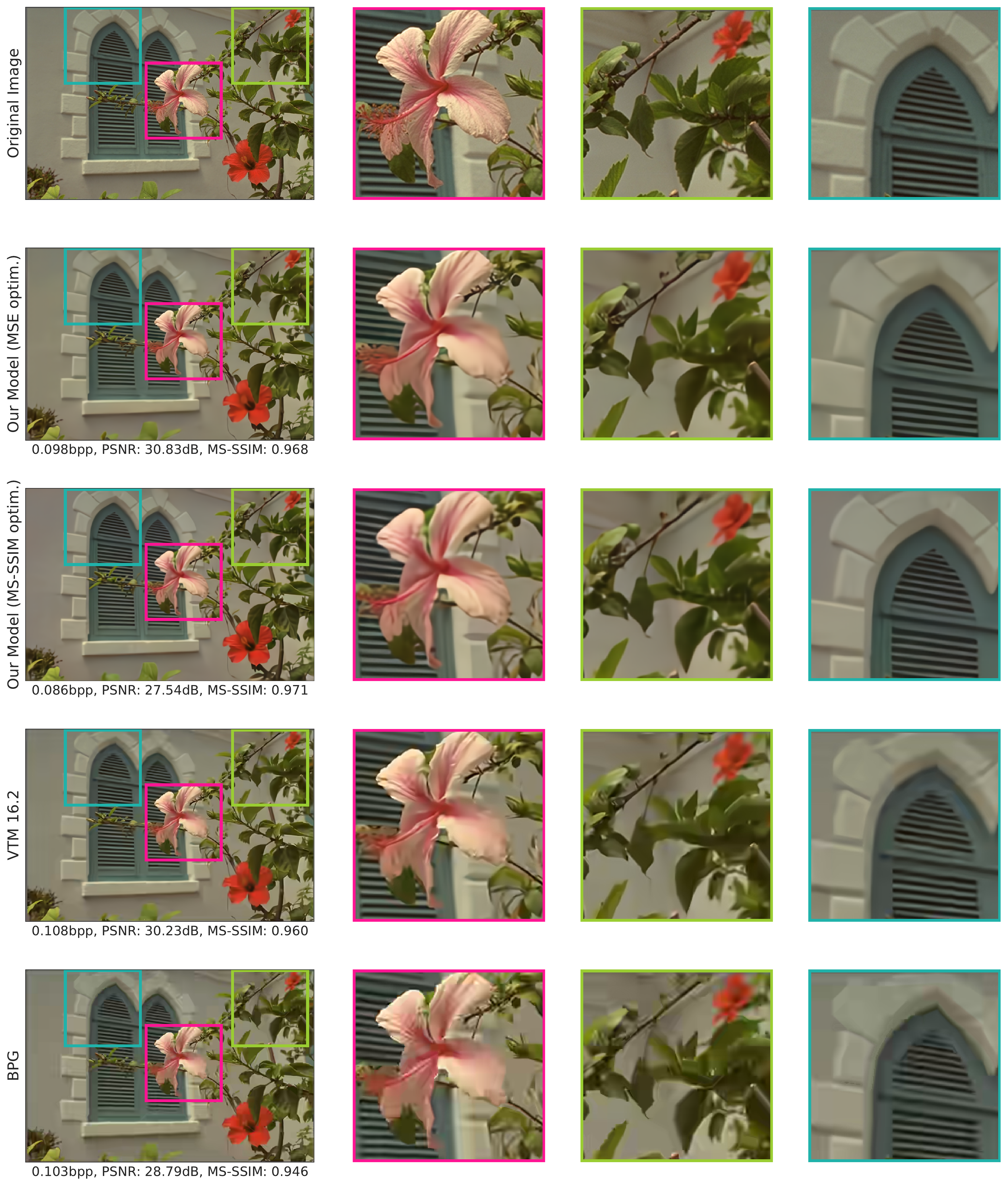}
	
	\caption{The reconstructed images of \textit{kodim07} from the Kodak image dataset for the visual comparison. The image is compressed by our models (optimized for MSE or MS-SSIM), VTM 16.2, and BPG for the target bpp of 0.1.}
	\label{fig:viz1}
\end{figure}

In order to assess qualitative performance, we visually compare the reconstructed images from our Contextformer models (base configuration), which are separately optimized for MSE and MS-SSIM, with the ones from BPG~\cite{bellard2015bpg} and VTM~16.2~\cite{jvet2019versatile}. \cref{fig:viz2} shows the reconstructed images of \textit{kodim23} (Kodak image dataset), and several crops from the images. Each reconstruction is generated by targetting approximately 0.06 bpp. Similarly, \cref{fig:viz1} shows the reconstructions of \textit{kodim07} for 0.1 bpp. As we can see from both figures that the classical codecs suffer from artifacts such as smear, blocking, and aliasing, whereas our methods (both MSE and MS-SSIM trained models) preserve contours and high frequency details better, and provides higher PSNR and MS-SSIM performance for lower bpp. While both the MSE and MS-SSIM trained models provide better visual quality than classical codecs, the type of distortions introduced by the Contextformer are influenced by the cost function used. In our experiments, the model optimized for MSE is better at producing sharper edges, while the model optimized for MS-SSIM is better at preserving structure and texture of the objects.



%
%

\end{document}